\documentclass[%
reprint,
amsmath,amssymb,
aps,
]{revtex4-2}

\usepackage{graphicx}
\usepackage{dcolumn}
\usepackage{bm}
\usepackage[colorlinks,linkcolor=blue,citecolor=blue,urlcolor=blue]{hyperref}
\usepackage{natbib}

\begin{document}

	\title{Tuning the anomalous Nernst and Hall effects with shifting the chemical potential in Fe-doped and Ni-doped Co\texorpdfstring{$_3$}{}Sn\texorpdfstring{$_2$}{}S\texorpdfstring{$_2$}{}}
	
	\author{Jie Liu$^{1}$, Linchao Ding$^{1}$,  Liangcai Xu$^{1}$, Xiaokang Li$^{1}$, Kamran Behnia$^{2}$,
		and Zengwei Zhu$^{1,\ddagger}$}
	
	\affiliation{$^1$Wuhan National High Magnetic Field Center and School of Physics, Huazhong University of Science and Technology,  Wuhan,  430074, China\\
		$^2$Laboratoire de Physique et Etude des Mat\'{e}riaux (CNRS/UPMC),Ecole Sup\'{e}rieure de Physique et de Chimie Industrielles, 10 Rue Vauquelin, 75005 Paris, France\\
	}
	
	\date{\today}
	
	\begin{abstract}
		Co$_3$Sn$_2$S$_2$ is believed to be a magnetic Weyl semimetal. It displays large anomalous Hall, Nernst and thermal Hall effects with a remarkably large anomalous Hall angle. Here, we present a comprehensive study of how substituting Co by Fe or Ni affects the electrical and thermoelectric transport. We find that doping alters the amplitude of the anomalous transverse coefficients.  The maximum decrease in the amplitude of the  low-temperature anomalous Hall conductivity $\sigma^A_{ij}$ is twofold. Comparing our results with theoretical calculations of the Berry spectrum assuming a rigid shift of the Fermi level, we find  that given the modest shift in the position of the chemical potential induced by doping, the experimentally observed variation occurs five times faster than expected. Doping affects the amplitude and the sign of the anomalous Nernst coefficient. Despite these drastic changes, the amplitude of the $\alpha^A_{ij}/\sigma^A_{ij}$ ratio  at the Curie temperature  remains close to $\approx 0.5 k_B/e$, in agreement with the scaling relationship observed across many topological magnets.

	\end{abstract}
	
	\maketitle
	
	\begin{figure}[t]
		\includegraphics[width=8cm]{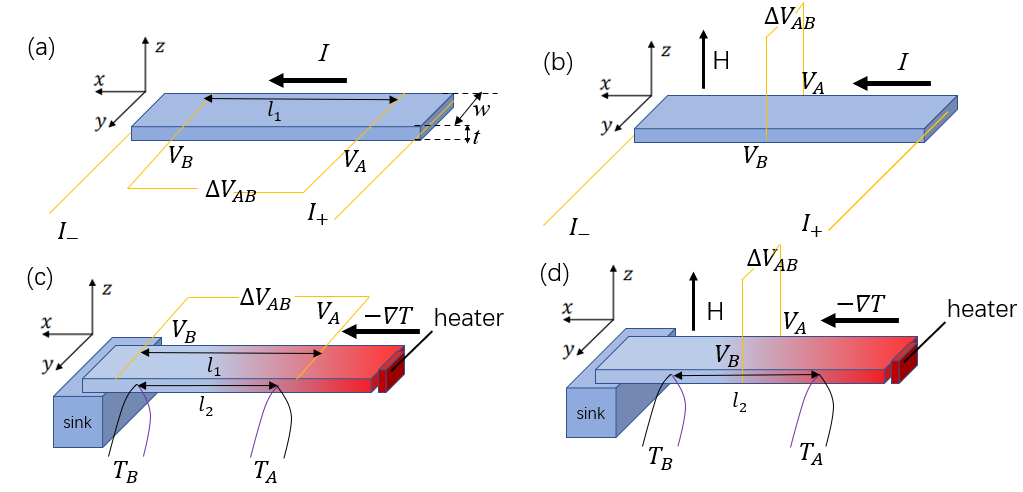}
		\caption{The experimental setup for measuring (a) longitudinal resistivity, (b) Hall resistivity, (c) Seebeck coefficient and (d) Nernst coefficient, respectively. The formulas to obtain the transport coefficients can be found in the section \uppercase\expandafter{\romannumeral2}.}
		\label{setup}
	\end{figure}
	
	\begin{figure}[t]
		\includegraphics[width=8.5cm]{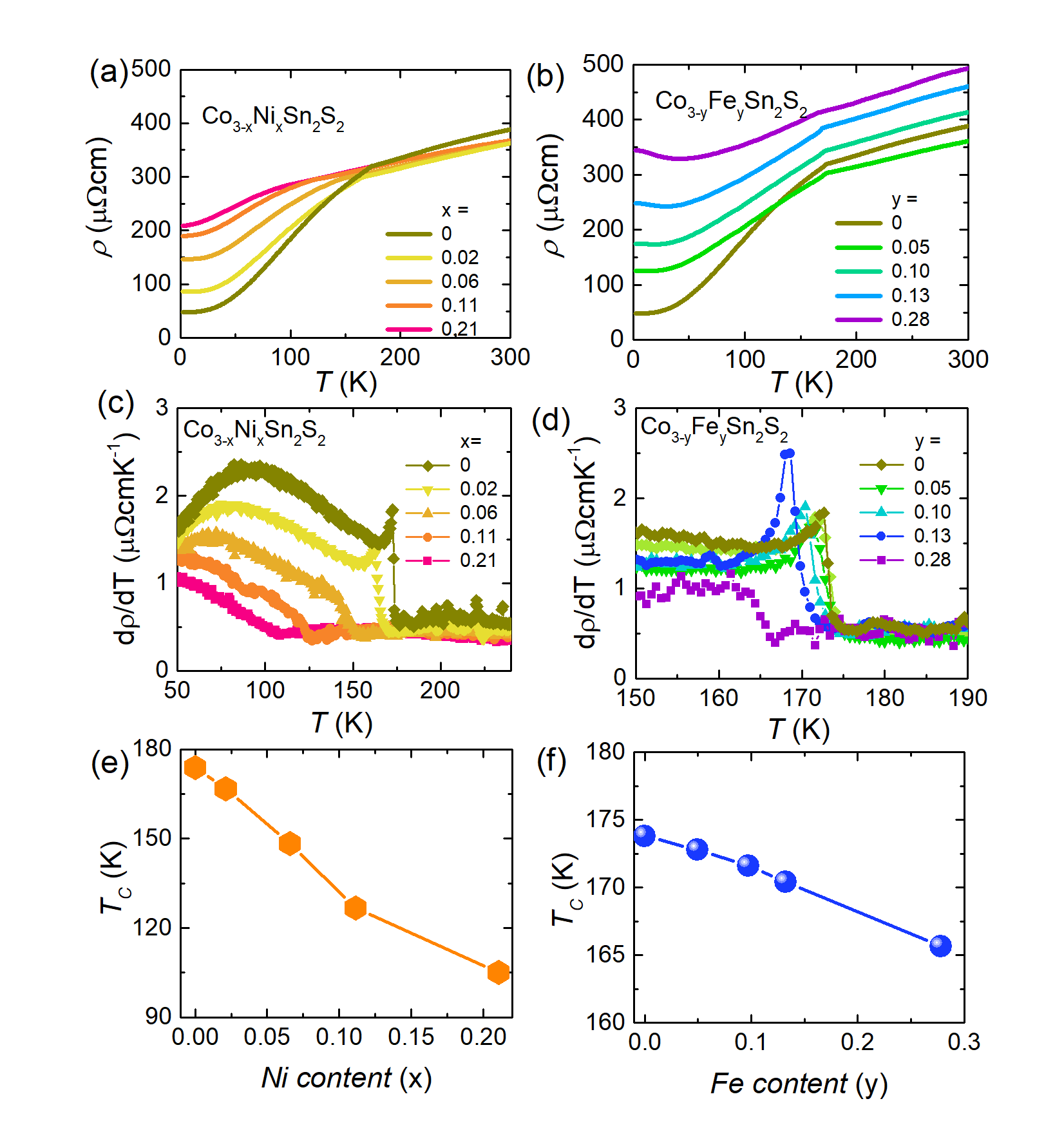}
		\caption{ (a) Temperature dependence of resistivity at zero-field of Ni doped Co$_3$Sn$_2$S$_2$ samples and (b) Fe doped samples. (c) Temperature dependence of the first derivative of resistivity for Ni doped Co$_3$Sn$_2$S$_2$ samples and (d) Fe doped samples. The evolution of Curie temperatures with doping ratio of  (e) Ni content $x$ and (f) Fe content $y$.}
		\label{rho_Sxx}
	\end{figure}

	\section{INTRODUCTION}
	
	In magnetic Weyl semimetals, the linear band structure on the Fermi surface gives rise to a significant local fictitious magnetic field in momentum space known as Berry curvature which is associated with the topological properties of the Bloch waves of electrons in the host solid \cite{Nagaosa2010, Xiao2010, Xiao2006}. The absence of time inversion symmetry leads to a nonzero integral of the Berry curvature in momentum space. The anomalous Hall effect (AHE) can have both intrinsic and extrinsic origins \cite{Nagaosa2010}. In the intrinsic case, this effect is attributed to the Berry curvature. The extrinsic mechanism refers to either skew scattering or `side jump' \cite{Nagaosa2010}. The anomalous Nernst effect (ANE) is a counterpart of the AHE. It emerges when mobile carriers respond to a temperature gradient  \cite{Onoda2008, Pu2008,Li2017PRL}. 
	
	The Mott relation \cite{Mott1971} states that the thermoelectric conductivity is set by the energy derivative of the electric conductivity at the Fermi level. This relation can be used to give an account of the measured Nernst signal in different contexts \cite{Behnia2015b}. A version of it is used to link the amplitude of the ordinary Nernst effect to the Hall mobility in metals \cite{Behnia2016}. Another version was invoked to give an account of a correlation observed between the amplitude of the anomalous Hall ($\sigma^A_{ij}$) and the anomalous Nernst ($\alpha^A_{ij}$) conductivities across different topological magnets \cite{Xu2020}. While the amplitude of each varies by orders of magnitude, their ratio at room temperature remains of the order of $k_B/e$ \cite{Xu2020}.

	The ANE is particularly sensitive to the band structure near Fermi level \cite{Xiao2006}. In 2D systems, the Fermi level can be tuned by varying the gate voltage. A sign reversal of ANE was revealed in the case of the magnetic topological insulator thin film
	Cr$_{0.15}$(Bi$_{0.1}$Sb$_{0.9}$)$_{1.85}$Te$_3$ by tuning the gate voltage \cite{Guo2017}. In 3D systems, an effective approach to tune the Fermi level is chemical doping, which can significantly influence the amplitude of the ANE. A number of systems have been studied in this way, including  Ga$_{1-x}$Mn$_x$As dilute magnets \cite{Pu2008} , La$_{1-x}$Sr$_x$CoO$_3$  ferromagnetic oxides  \cite{Miyasato2007}, as well as 
	${\mathrm{Co}}_{2}{\mathrm{MnAl}}_{x}{\mathrm{Si}}_{1\ensuremath{-}x}$ magnetic Weyl metals \cite{Breidenbach2022}.

	The Shandite compound Co$_3$Sn$_2$S$_2$ is a magnetic Weyl semimetal with a remarkably large intrinsic anomalous Hall conductivity (AHC) (up to 1130 $\Omega^{-1}$cm$^{-1}$) and anomalous Hall angle (up to 20\% \cite{Liu2018CSS}). Its Curie temperature is 177 K and the magnetic moment corresponds to 0.29 $\mu_B$/Co. Calculations and experiments indicated the presence of Weyl points located at 60 meV above the Fermi level \cite{Wang2018,Liu2018CSS,Shen2020PRL,Muechler2020}. This allows to adjust the topological property easily through doping \cite{Shen2020,Corps2015,Zhou2020PRB,Thakur2020,Yanagi2021} or by applying pressure \cite{Chen2019PRB,Liu2020PRM,Zeng2022}. The cobalt site can be doped by either nickel or iron. The tin site can be occupied by indium. Upon chemical doping by iron, the AHC is further enhanced to 1850 $\Omega^{-1}$cm$^{-1}$ and anomalous Hall angle up to 33\% \cite{Shen2020}. The elevations of AHC ($\sim1400~\Omega^{-1}$cm$^{-1}$), AHA($\sim$22\%) in Ni-doped samples was attributed to the local disorder effect broadening the inverted bands \cite{Shen2020PRL}, which remains intrinsic \cite{Thakur2020}. Indium doping enhances AHC \cite{Zhou2020PRB} and introduces additional quantum phase transitions to the phase diagram \cite{Guguchia2021,Fujioka2014,Yanagi2021}. A successful attempt in  introducing electrons by replacing S with Sb has also been reported \cite{Li2022Sb}. 
	
	The anomalous Nernst effect in pristine Co$_3$Sn$_2$S$_2$ has been the subject of numerous studies \cite{Yuke2020,Guin2019,Ding2019}. By studying samples of different mobilities, Ding \textit{et al.} \cite{Ding2019} found that the ordinary (anomalous) Nernst effect (S$_{yx}^{O}$) increases (decreases) with increasing carrier mobility. This anti-correlation between ANE (S$^A_{yx}$) and carrier mobility was found to be in agreement with the intrinsic origin of the anomalous Nernst conductivity ($\alpha^A_{yx}$) set by Berry curvature . On the theoretical side, calculations predict that In doping largely impacts the ANE \cite{Yanagi2021}. 
	
	Here, we present a systematic study of AHE and ANE in  Co$_3$Sn$_2$S$_2$ samples doped with Fe or Ni. We find that the amplitude and the temperature dependence of AHE and ANE display a gradual evolution. The  amplitude of the low-temperature Hall-conductivity $\sigma^A_{ij}$ decreases by a factor of 2. The low-temperature anomalous Nernst effect changes sign. The variation of $\sigma^A_{ij}$ with doping is much more rapid that what is expected by theoretical expectations based on the quantification of the Berry spectrum \cite{Ding2019} neglecting correlations and extrinsic contributions. Despite the notable variation in the temperature dependence of the Hall and Nernst conductivities,  $\alpha^A_{ij}/\sigma^A_{ij}$ ratio barely changes at the Curie temperature. In doped samples, it displays an almost linear temperature dependence and  and becomes $ \approx 0.5 k_B/e$ near  $T_{\rm{C}}$, in agreement with the scaling relation found in other topological magnets \cite{Xu2020}.
	
	\section{METHODS}
	
	Electron doping  is achieved  by replacing Co with Ni and hole doing by substituting Co with Fe \cite{Kassem2016}. A stoichiometric ratio of Ni, Fe, Co, Sn and S powders was sealed in a quartz tube. We grew high quality single crystals of Co$_{3-x}$Ni$_x$Sn$_2$S$_2$ and Co$_{3-y}$Fe$_y$Sn$_2$S$_2$ by self-flux method as reported before \cite{Ding2019,Ding_2021}.  
	
	Shiny crystals were selected by careful mechanical separation after growth. Crystal stoichiometry was confirmed by energy dispersive X-ray spectroscopy (EDS) and the results are shown in Table \ref{Table1}. Transport measurements were performed with a physical property measurement system (Quantum Design PPMS-9). A thermal or electrical current was applied along the $a$-axis \cite{Ding2019}. Longitudinal and Hall resistivity were measured by a standard four-probe method. The Seebeck  and the Nernst coefficients were measured using a chip-resistance heater and two pairs of thermocouples under high-vacuum environment of the PPMS. The schematic illustration of the experimental setup is shown in Fig. \ref{setup}. We can obtain longitudinal resistivity, Hall resistivity, Seebeck coefficient and Nernst coefficient by $\rho_{xx}=\frac{\Delta V_{AB}wt}{Il_1}$, $\rho_{yx}=\frac{\Delta V_{AB}t}{I}$, $S_{xx}=\frac{E_x}{\nabla_x T}=-\frac{\Delta V_{AB}/l_1}{\Delta T_{AB}/l_2}$, $S_{yx}=\frac{E_y}{\nabla_x T}=-\frac{\Delta V_{AB}/w}{\Delta T_{AB}/l_2}$, respectively. Where $\Delta V_{AB}=V_{A}-V_{B}$, $\Delta T_{AB}=T_{A}-T_{B}$.

	\begin{table}[b]
		\centering
		\caption{Elemental composition analysis results from the energy dispersive X-ray spectroscopy (EDS) for Co$_{3-x}$Ni$_x$Sn$_2$S$_2$ and Co$_{3-y}$Fe$_y$Sn$_2$S$_2$  single crystal samples with different doping ratios.}
		\label{Table1}
		\begin{tabular}{ccccc|cccccc}
			\multicolumn{5}{c}{Co$_{3-x}$Ni$_x$Sn$_2$S$_2$} & \multicolumn{5}{c}{Co$_{3-y}$Fe$_y$Sn$_2$S$_2$ }\\
			\hline
			$x$  &Co    &Ni    & Sn   & S    & $y$  &Co    &Fe    & Sn   & S       \\
			\hline
			0.02 & 2.98 & 0.02 & 2.07 & 1.92 & 0.05 & 2.95 & 0.05 & 2.09 & 1.91 \\
			0.06 & 2.94 & 0.06 & 2.11 & 1.91 & 0.10 & 2.90 & 0.10 & 2.08 & 1.90 \\
			0.11 & 2.89 & 0.11 & 2.09 & 1.93 & 0.13 & 2.87 & 0.13 & 2.09 & 1.90 \\
			0.21 & 2.79 & 0.21 & 2.10 & 1.87 & 0.28 & 2.72 & 0.28 & 2.05 & 1.87 \\ 
			\hline
		\end{tabular}
		
	\end{table}

	\section{RESULTS AND DISCUSSION}
	
	\begin{figure}[t]
		\includegraphics[width=8cm]{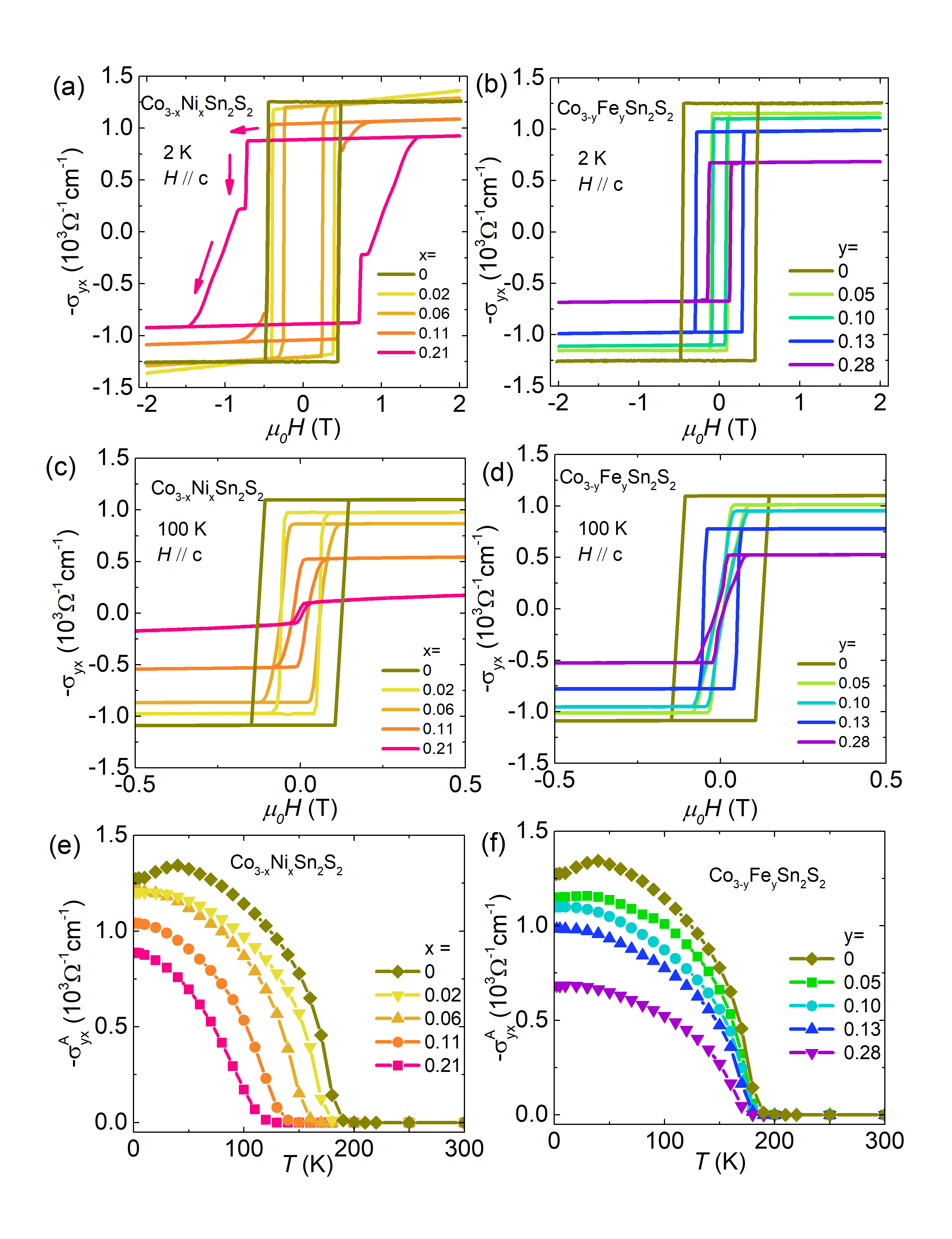}
		\caption{(a). Field dependence of the Hall conductivity $\sigma_{yx}$ of Co$_{3-x}$Ni$_x$Sn$_2$S$_2$ and (b). Co$_{3-y}$Fe$_y$Sn$_2$S$_2$ single crystal samples with different doping ratios at 2 K, respectively. (c). Field dependence of the Hall conductivity $\sigma_{yx}$ of Co$_{3-x}$Ni$_x$Sn$_2$S$_2$ and (d). Co$_{3-y}$Fe$_y$Sn$_2$S$_2$ at 100 K. (e). Temperature dependence of anomalous Hall conductivity $\sigma^A_{yx}$ of Co$_{3-x}$Ni$_x$Sn$_2$S$_2$ and (f). Co$_{3-y}$Fe$_y$Sn$_2$S$_2$.}
		\label{Hall}
	\end{figure}
	
	\begin{figure}
		\includegraphics[width=8.5cm]{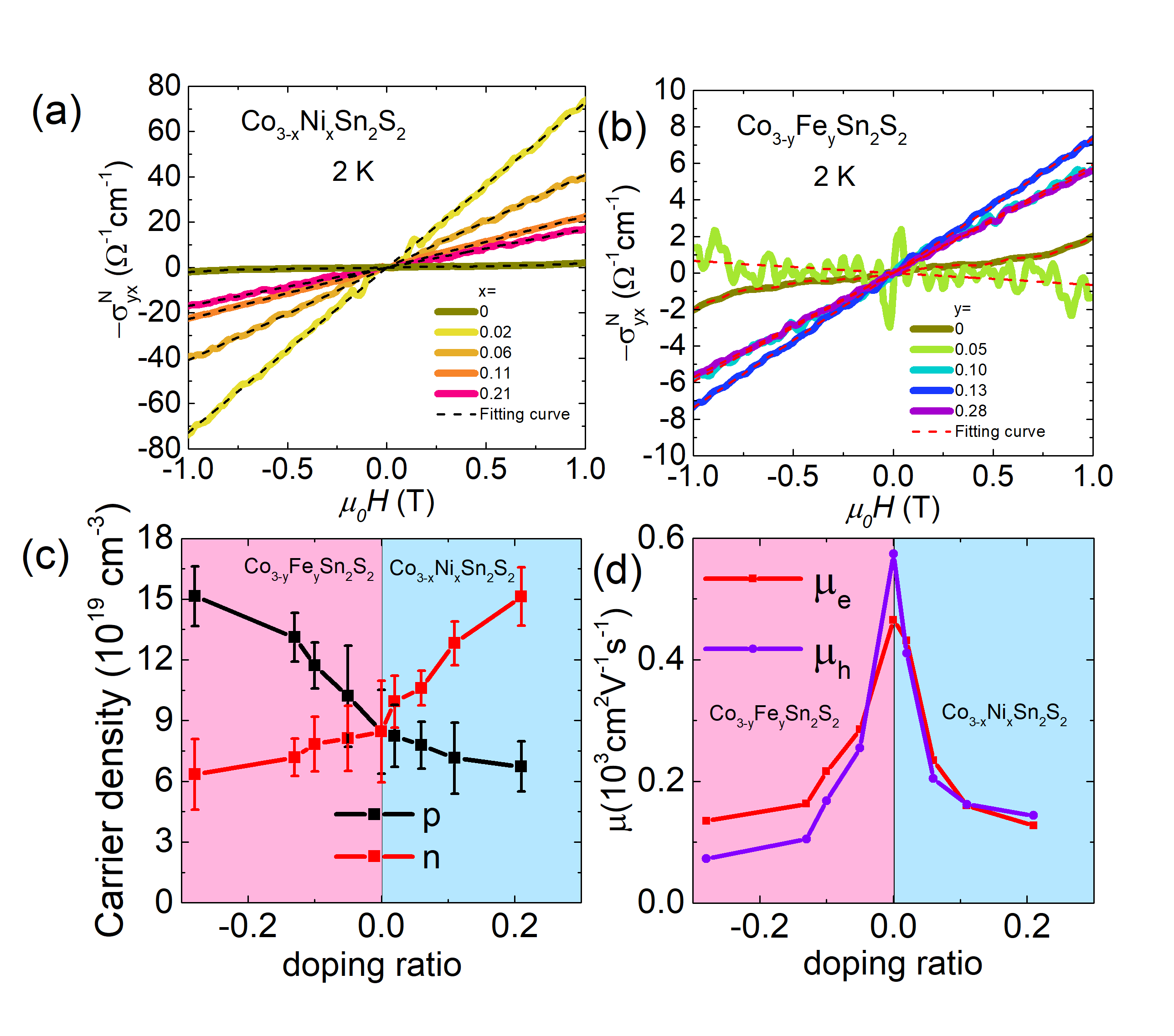}
		\caption{(a). Low-field ordinary Hall conductivities of Ni doped Co$_3$Sn$_2$S$_2$ samples and (b) Fe doped samples. Dashed lines are the fits with the two-band model. (c) The doping ratio dependence of carrier density. (d) Doping ratio dependence of mobilities obtained from the two-band model. The moiblities are reduced as dopants are increased. }
		\label{twoband}
	\end{figure}

	\begin{figure}
		\includegraphics[width=8.5cm]{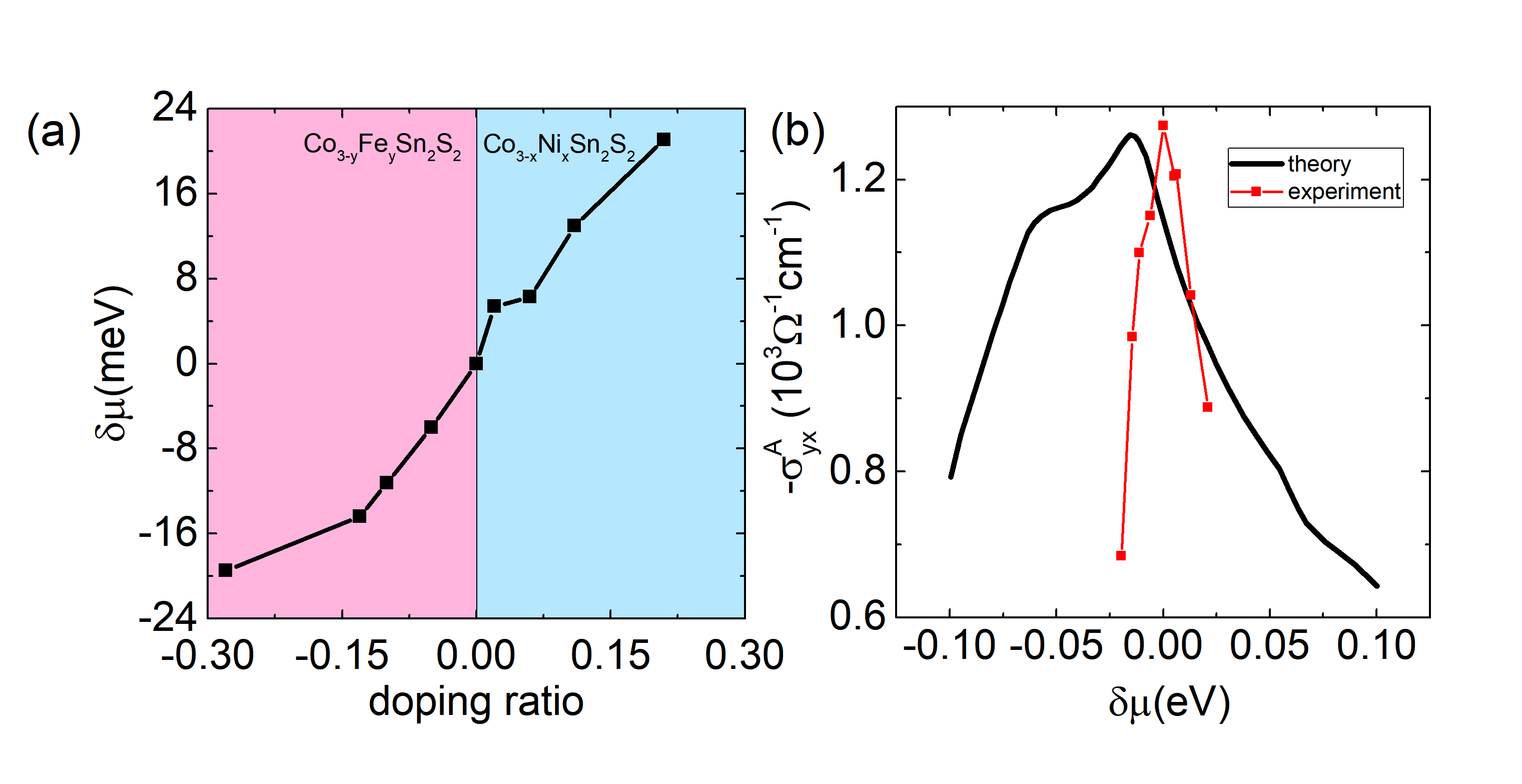}
		\caption{(a) The doping ratio dependence of the shifted chemical potential. (b)The chemical potential dependence of the anomalous Hall conductivity. The black curve represents the result of theory\cite{Ding2019} for the intrinsic contribution from Berry curvature, while the red curve is the experimental measured AHE result.}
		\label{sigma_chemical}
	\end{figure}
	
	\begin{table}[b]
		\centering
		\caption{Electron (hole) carrier density [$n (p)$] (in units of $\mathrm{10^{19} cm^{-3}}$) together with their mobilities [$\mu_e (\mu_h)$] (in units of $\mathrm{cm^2 V^{-1} s^{-1}}$) in different samples. They were extracted using a two-band fit of the ordinary Hall conductivity.}
		\label{Table2}
		\begin{tabular}{cccccc|cccccc}
			\multicolumn{6}{c}{Co$_{3-x}$Ni$_x$Sn$_2$S$_2$} & \multicolumn{5}{c}{Co$_{3-y}$Fe$_y$Sn$_2$S$_2$ }\\
			\hline
			$x$ & 0 &0.02   &0.06   & 0.11  & 0.21   & $y$  &0.05   &0.10   & 0.13  & 0.28      \\
			\hline
			p & 8.45 & 8.24 & 7.78 & 7.14 & 6.73 &  & 10.20 & 11.72 & 13.11 & 15.14 \\
			n & 8.46 & 9.93 & 10.60 & 12.81 & 15.12 &  & 8.12 & 7.83 & 7.18 & 6.34 \\
			$\mu_h$ & 574 & 411 & 205 & 162 & 144 &  & 255 & 168 & 105 & 73 \\
			$\mu_e$ & 465 & 431 & 234 & 160 & 127 &  & 285 & 216 & 163 & 135 \\ 
			\hline
		\end{tabular}
		
	\end{table}
	
	\begin{figure*}[t]
		\includegraphics[width=18cm]{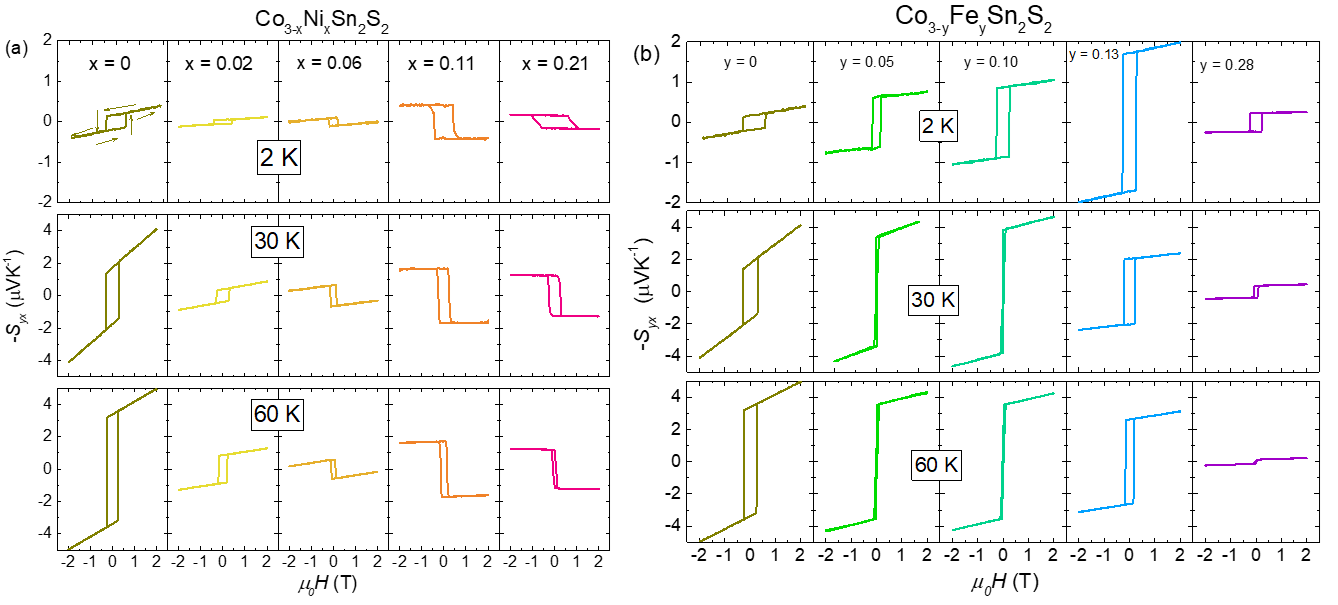}
		\caption{ 
			(a) The field dependence of Nernst effect $S_{yx}$ of Co$_{3-x}$Ni$_x$Sn$_2$S$_2$ and (b) Co$_{3-y}$Fe$_y$Sn$_2$S$_2$ at three typical temperatures: 2 K, 30 K and 60 K. }
		\label{syx}
	\end{figure*}

	Fig. \ref{rho_Sxx}(a) and \ref{rho_Sxx}(b) show the zero-field temperature-dependent resistivity of Co$_{3-x}$Ni$_x$Sn$_2$S$_2$ and Co$_{3-y}$Fe$_y$Sn$_2$S$_2$ samples. The Ni-substituted samples remain metallic with a residual resistivity, which increases with increasing $x$. In the case of Fe-substituted samples, the resistivity shows a low-temperature upturn when $y$ exceeds  0.1 as reported previously  and attributed to a Kondo effect \cite{Shen2020}. Ordinarily, the Kondo effect completes with ferromagnetism \cite{Pasupathy2004}. However, the Kondo correlations can still be present, even ferromagnetism intends to suppress Kondo-assisted tunneling \cite{Pasupathy2004}. Several compounds has been proposed as the ferromagnetic Kondo system \cite{Krellner2007,Tursina2018,Das2019}. Interestingly, the Fe-doped samples with similar contents  evolves earlier into the Kondo-like state. This may be due to the relative strong magnetism in iron. The coexist of ferromagnetism and Kondo effect would need to be further clarified.  Since $k_f \ell = 44$ \cite{Ding_2021}  when the residual resistivity is 200 $\mu \Omega$cm, one can exclude Anderson localozation. 
	
	At Curie temperature, there is a visible kink in the temperature dependence of resistivity. In both Ni and Fe doped samples, this kink shifts to lower temperature with the increase in the dopant concentration. This is better illustrated in the temperature dependence of the first derivative of resistivity shown in Fig. \ref{rho_Sxx}(c) and Fig. \ref{rho_Sxx}(d). Fig. \ref{rho_Sxx}(e) and Fig. \ref{rho_Sxx}(f) show the evolution of the Curie temperature with doping contents, $x$ and $y$. In Co$_{3-x}$Ni$_x$Sn$_2$S$_2$ samples, the Curie temperature decreases with increasing $x$ and the variation is very close to the results reported before \cite{Thakur2020,Shen2020}. However, in Co$_{3-y}$Fe$_y$Sn$_2$S$_2$, a doping content of $y = 0.28$ pulls down the Curie temperature by less than 10 K. This is inconsistent with a previous report where the Curie temperature decreased by 25 K with a doping content of 0.2 \cite{Shen2020}. The discrepancy is probably due to the different flux growth methods employed. Shen \textit{et al.} \cite{Shen2020} used Sn and Pb mixed flux to grow samples. On the other hand, our samples were grown by self-flux method. This calls for further studies for clarification.
	
	The temperature dependence of Hall conductivity can be extracted from longitudinal resistivity $\rho_{xx}$ and Hall resistivity $\rho_{yx}$, using  $\sigma_{yx}=\frac{-\rho_{yx}}{\rho_{xx}^2+\rho_{yx}^2}$. Fig. \ref{Hall}(a) and \ref{Hall}(b) present the magnetic field dependence of the Hall conductivity in Co$_{3-x}$Ni$_x$Sn$_2$S$_2$ and  Co$_{3-y}$Fe$_y$Sn$_2$S$_2$ samples at 2 K with the field oriented along the $c$-axis. In Co$_{3-x}$Ni$_x$Sn$_2$S$_2$, the Hall conductivity shows an obvious hysteresis. The behavior changes as the doping increases. When the doping level is low, the jump in the Hall conductivity is very steep. After $x > 0.06$, the jump begins to widen. When the doping ratio is $x =0.21$, the jump occurs in two steps along the direction indicated by the arrows in the figure. 
	
	A similar process was reported in ref. \cite{Shen2020PRL,Thakur2020}. The two-step jumps in the field dependence of the magnetization observed near the Curie temperature indicates mixed magnetic phases with  magnetic moments oriented along in-plane and out-of-plane directions \cite{Shen2020PRL,Thakur2020}. When the magnetic field is swept, the inconsistency between the moment reversal process along in-plane and out-of-plane orientations leads to a two-step jump \cite{Kassem2021}. With the increase of doping ratio $x$, the mixed magnetic phases survive down to low temperatures, indicating that Ni doping affects the magnetic structure. In contrast, Fe-doped samples display sharp jumps in the field dependence of their magnetization at low temperatures.  Two-step jumps reappear when temperature becomes as high as 100 K. Both types of doping reduce the coercive field and the amplitude of the anomalous Hall signal, as shown in Fig. \ref{Hall}(c) and \ref{Hall}(d). The detailed magnetic structures of the doped samples is a subject for future experiments.
	
	Fig. \ref{Hall}(e) and Fig. \ref{Hall}(f) show the temperature dependence of the AHC.  With the increase of temperature, the magnetic moment tends to be disordered with the increase of thermal disturbance. Above the Curie temperature, the Co$_3$Sn$_2$S$_2$ is no more a magnetic Weyl semimetal, thus AHC tends to be 0 \cite{Shen2020}. As seen in Fig. \ref{Hall}(e), AHC at 2 K decreases with the increase in the doping level in Ni-substituted samples. This is consistent with what was reported for Ni-substituted samples grown by self-flux method \cite{Thakur2020}, but not with Ni-substituted samples grown by flux method by  Shen \textit{et al.} who found that AHC was enhanced to 1400 $\Omega^{-1}$cm$^{-1}$ when x=0.1 \cite{Shen2020PRL}. The situation is similar in the case of Fe-substituted samples. As seen in Fig. \ref{Hall}(f), in Co$_{3-y}$Fe$_y$Sn$_2$S$_2$ grown by self-flux method, the 2 K AHC decreases monotonically with increasing $y$. In the Fe-substituted samples obtained by Shen \textit{et al.}  using the Sn and Pb flux method, AHC increased to 1800 $\varOmega^{-1}$cm$^{-1}$ when y=0.05 \cite{Shen2020} in sharp contrast with what is observed here in  samples obtained by the self-flux method here. This difference indicates that different growth methods influence the band structure and possibly the lattice constant.

	\begin{figure*}[t]
		\includegraphics[width=18cm]{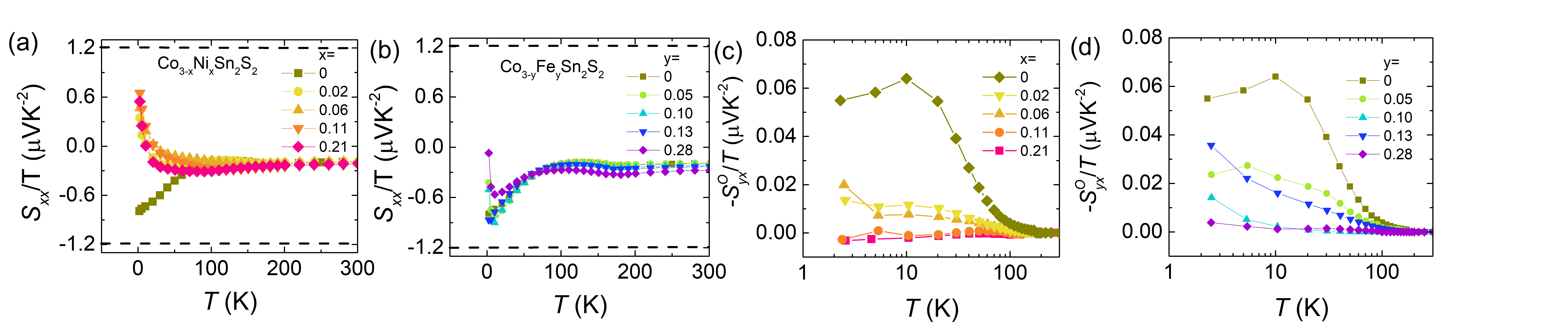}
		\caption{(a) Temperature dependence of the Seebeck coefficient over temperature at zero-field of Ni doped Co$_3$Sn$_2$S$_2$ samples and (b) of Fe doped samples. (c). Temperature dependence of $S^{O}_{yx}/T$ in samples with different doping ratios of Co$_{3-x}$Ni$_x$Sn$_2$S$_2$ and (d) Co$_{3-y}$Fe$_y$Sn$_2$S$_2$, respectively.}
		\label{ordinary}
	\end{figure*}
	
	\begin{figure*}[t]
		\includegraphics[width=18cm]{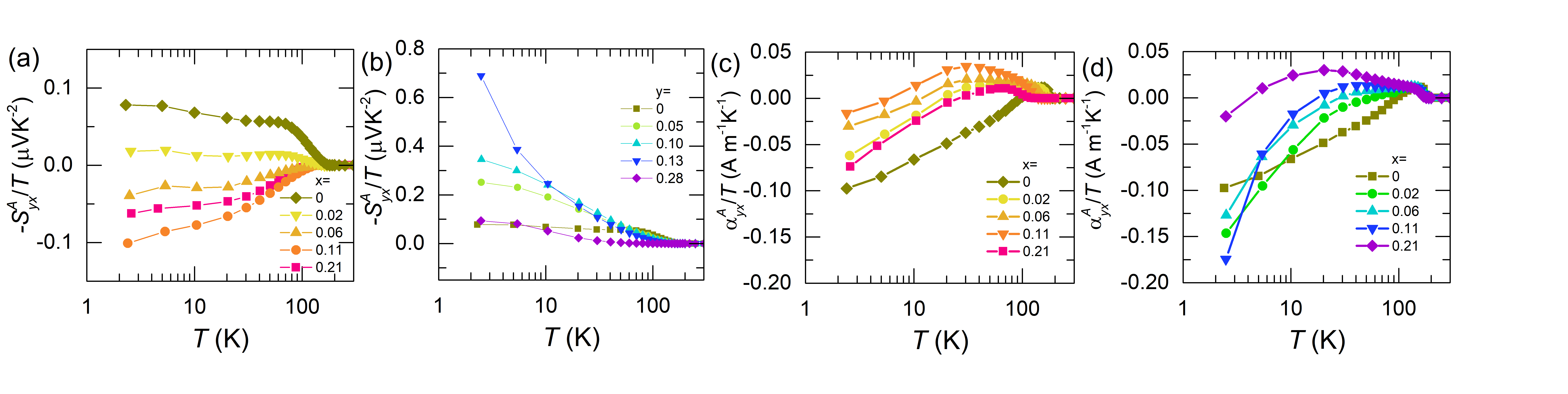}
		\caption{(a). Temperature dependence of $S^{A}_{yx}/T$ for Co$_{3-x}$Ni$_x$Sn$_2$S$_2$ and (b) Co$_{3-y}$Fe$_y$Sn$_2$S$_2$, respectively.  (c) Temperature dependence of anomalous off-diagonal thermoelectric conductivity $\alpha^A_{yx}/T$ in Co$_{3-x}$Ni$_x$Sn$_2$S$_2$ and (d) Co$_{3-y}$Fe$_y$Sn$_2$S$_2$, respectively.}
		\label{anomalous}
	\end{figure*}
	
	We used a two-band model to fit the field dependence of the Hall conductivity and to quantify the evolution of carrier density and mobility with doping. Fig. \ref{twoband} (a) and (b) present fits to the extracted ordinary Hall conductivity at 2K, after subtracting the AHC component. For accuracy, the extracted parameters were checked with zero-field resistance. The parameters obtained from the fitting are listed in table \ref{Table2}. Fig. \ref{twoband} (c) displays the doping dependence of the carrier density of electrons ($n$) and holes ($p$). The electron carrier density ($n$) exceeds that of holes ($p$) in Ni-doped samples. In Fe-doped samples, the tendency is opposite. As expected, the carrier mobility decreases with both Co and Ni doping as shown in the Fig. \ref{twoband} (d). 
	
	In order to compare with theory, we estimated the shift of the chemical potential induced by doping. It is shown in Fig. \ref{sigma_chemical} (a). It was calculated using this equation:
	\begin{equation}
		\delta\mu=\frac{\hbar^2\pi^2(3\pi^2n_{ave})^{-1/3}}{2m^*}\delta n+\frac{\hbar^2\pi^2(3\pi^2n_{ave})^{-1/3}}{2m^*}\delta p 
		\label{1}
	\end{equation}
	
	We took  $m^*$=0.65$m_0$\cite{Ding_2021}, as effective mass, which is the average value obtained from quantum oscillations in the undoped sample. The average carrier density was taken to be: $n_{ave}=\frac{p+n}{2}$.
	
	Neglecting any variation of the effective mass $m^*$ with doping, one finds that the chemical potential varies from -19 meV to 20 meV for maximum $p$- and $n$-doping. Fig. \ref{sigma_chemical} (b) compares the variation of the experimentally-measured and theoretically computed AHC. They resemble qualitatively, but not quantitatively.  There is a five-fold difference in the horizontal scale.  Let us recall that there is also a twofold discrepancy between theoretically calculated and experimentally measured carrier density \cite{Ding_2021}, presumably because theory neglects electron-electron interaction \cite{Ding_2021}. The fivefold discrepancy observed here may be partially due to a twofold mass renormalization due to interaction. Taken at its face value, it points to the existence of an extrinsic contribution with a  sign of  opposite to the intrinsic one.

	\begin{figure*}[t]
		\includegraphics[width=16cm]{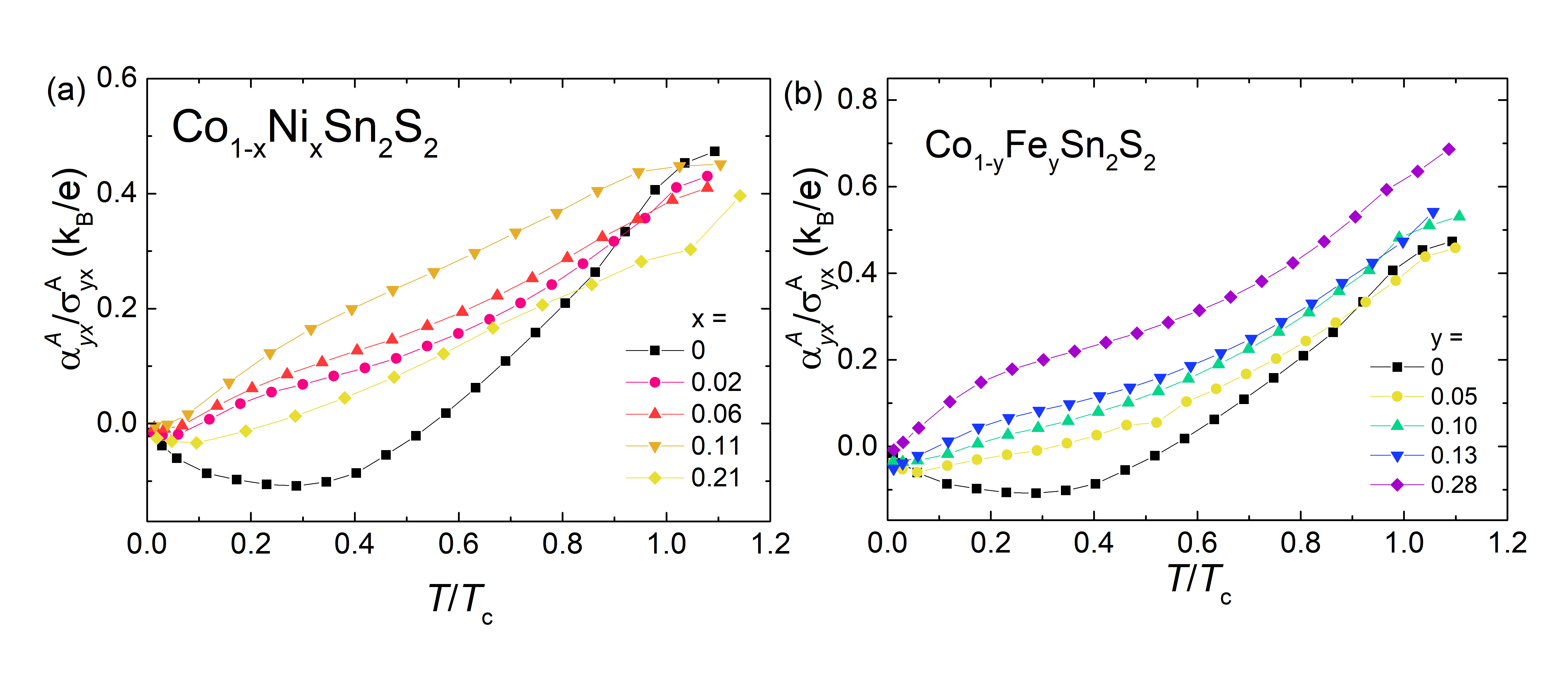}
		\caption{(a) The temperature dependence of the ratio for $\alpha^A_{ij}/\sigma^A_{ij}$ of Co$_{3-x}$Ni$_x$Sn$_2$S$_2$ and  (b) for Co$_{3-y}$Fe$_y$Sn$_2$S$_2$, respectively. The $\alpha^A_{ij}/\sigma^A_{ij}$ remains close to $\approx $ 0.5 $k_b/e$ near the Curie temperature for all samples. }
		\label{ratio_T}
	\end{figure*}

	Fig. \ref{syx} shows the field dependence of Nernst response for all doped samples at 2 K, 30 K and 60 K respectively. The Nernst signal consists of a slope (caused by the ordinary component) and a jump with hysteresis (caused by the anomalous components). In Co$_{3-x}$Ni$_x$Sn$_2$S$_2$, the sign of the anomalous Nernst signal $S^{A}_{yx}$ changes from positive to negative between $x=0.02$ to $x=0.06$. In Co$_{3-y}$Fe$_y$Sn$_2$S$_2$, the ANE enhances with increasing $y$, and attains a maximum at $y=0.13$. Afterwards, it decreases. The largest amplitude observed ($S^{A}_{yx}\approx -4 \mu V/K$) was for a doping of $y=0.10$ at 30 K.
	
	Fig. \ref{ordinary} shows the temperature dependence of the Seebeck coefficient and the ordinary Nernst effect in all samples. $S_{xx}/T$ for Co$_{3-x}$Ni$_x$Sn$_2$S$_2$ is shown in Fig. \ref{ordinary}(a) and for Co$_{3-y}$Fe$_y$Sn$_2$S$_2$ in Fig. \ref{ordinary}(b). In the pristine sample, $S_{xx}/T$ is $\approx -0.2 \mu V/K^2$ in most of the temperature range and then at below 100 K it tends towards $-0.8 \mu V/K^2$. Co$_3$Sn$_2$S$_2$ is a compensated semi-metal with  numerous small electron and hole pockets \cite{Ding_2021}. Because of the contributions of opposite signs, its overall Seebeck response should be smaller than what is  expected $S_{xx}/T$ from the Fermi temperature in a one-band picture \cite{Kamran2004}:  $S_{xx}/T=\frac{\pi^2}{3}\frac{k_B}{e}\frac{k_B}{E_F}=|1.2|$ $\mu$V/K$^2$. The average Fermi energy of a pocket can be estimated to be $\approx 20 meV$ using $E_F=\frac{\pi^2n_{ave}k_B^2}{2\gamma}$.  Here  $\gamma=1.59$ mJ/(molK$^2$) \cite{Hu2022specific} is the Sommerfeld coefficient and $n= 8.8\times 10^{25}cm^{-3}$ \cite{Ding2019} is the carrier density. If the only carriers were electrons (holes), the expected $S_{xx}/T$ would have been $- (+)1.2 \mu$V/K$^2$. In the perfectly compensated Co$_3$Sn$_2$S$_2$, $S_{xx}/T$ should be somewhere between depending on the mobility of the hole-like and electron-like carriers.  The fact that it is negative at low temperature indicates that that the mean-free-path of occupied states is longer. In the Ni-doped samples (Fig. \ref{ordinary}(a)) $S_{xx}/T$ becomes positive below 20 K, which indicates that the mean-free-path of unoccupied states (`hole-like' carriers) becomes longer than the occupied states ('electron-like' carriers). In the case of Fe-doped samples, doping causes little change in $S_{xx}/T$, apart from a slight increase at temperatures. Note that the sign of the Seebeck coefficient depends on the competing contributions by occupied and unoccupied states. It is positive in copper, despite an electron-like Fermi surface, because unoccupied states contribute more than occupied states to the overall Seebeck response \cite{Behnia2015b,Robinson1967}.

	The temperature dependence of the ordinary Nernst effect (ONE) extracted from our data is  shown in Fig. \ref{ordinary} (c)-(d). As seen in Fig. \ref{ordinary} (c), in Co$_{3-x}$Ni$_x$Sn$_2$S$_2$, the ordinary Nernst coefficient $S^{O}_{yx}/T$ decreases rapidly with doping, which is expected given the reduction in the mean free path \cite{Kassem2016}. At $x=0.2$,  ONE has a sign change. Fe doping also leads to a reduction in ONE, as seen in Fig. \ref{ordinary}(d). The comparison shows that ONE is less affected by Fe doping than by Ni doping. 
	
	Fig. \ref{anomalous} shows the ANE and the anomalous off-diagonal thermoelectric conductivity $\alpha_{yx}^A$ in all samples. As seen in Fig. \ref{anomalous}(a), doping induces a complex evolution of the ANE. The sign of  $S^{A}_{yx}$ changes at all temperatures for $x>0.06$. In Fig. \ref{anomalous} (b), one can see that compared to Ni doping,  Fe doping has a larger impact on the low-temperature ANE. In addition, Fe doping also greatly enhances the anomalous Nernst coefficient in the low temperature region, up to six times.  
	
	The sign change of anomalous Nernst coefficient in Fig. \ref{anomalous}(a) has been observed in Co$_{3-x}$Ni$_x$Sn$_2$S$_2$ as the doping ratio is increased, while  the Nernst signal keeps its positive sign with Fe doping in the whole doping range. Such a change induced by the shift of Fermi level, is not accompanied with a sign change in AHC. In the dilute magnetic semiconductor material Ga$_{1-x}$Mn$_x$As,  the anomalous Nernst coefficient changes sign with doping \cite{Pu2008}.  In the magnetic topological insulator thin film Cr$_{0.15}$(Bi$_{0.1}$Sb$_{0.9}$)$_{1.85}$Te$_3$, the Fermi level can be tuned by the gate voltage $V_g$ \cite{Guo2017}. When  $V_g >100$ V, the anomalous Nernst effect changes sign below 5K. Fig. \ref{anomalous}(c) and Fig. \ref{anomalous}(d) shows the temperature dependence of anomalous off-diagonal thermoelectric conductivity $\alpha^A_{yx}/T$ with temperature which is calculated from $\alpha^A_{yx}=S_{xx}\sigma^A_{yx}+S^A_{yx}\sigma_{xx}$ \cite{Ding2019}. One can see that $\alpha^A_{yx}/T$ displays a large variation below the Curie temperature in both Co$_{3-x}$Ni$_x$Sn$_2$S$_2$ and Co$_{3-y}$Fe$_y$Sn$_2$S$_2$. $\alpha^A_{xy}$ is determined by the Berry curvature near the Fermi level. The energy band structure obtained by theoretical calculation shows that the Berry curvature exhibits dominantly negative value near the charge neutral point, which is caused by anticrossing bands \cite{Ding2019}. Then large negative Berry curvature induces large negative $\alpha^A_{xy}/T$ at low temperature. Thus, the Berry curvature determines the sign change of $\alpha^A_{xy}$.

	It is instructive to plot the  $\alpha^A_{ij}$ and $\sigma^A_{ij}$ ratio. Despite the complex evolution of both, this ratio  does not show a strong variation at high temperature. As shown in the Fig. \ref{ratio_T}, the $\alpha_{yx}^A/\sigma_{yx}^A$ ratio remains close to 0.5 k$_B$/e near the Curie temperature and smoothly vanishes towards zero with decreasing temperature.  It was already observed \cite{Xu2020,asaba2021}, that the amplitude of  $\alpha^A_{ij}$/$\sigma^A_{ij}$ in many topological magnetic materials is of the same order of magnitude. In the intrinsic picture, such a correlation arises because both quantities average the Berry curvature of the Fermi surface (but with pondering factors) \cite{Ding2019,Xu2020}. Our observation that the $\alpha_{yx}^A/\sigma_{yx}^A$ ratio barely changes by doping suggests that even in presence of a sizeable extrinsic contribution the $\alpha^A_{ij}$/$\sigma^A_{ij}$ ratio remains a sizable fraction of $k_B/e$ near the ordering temperature.

	\section{CONCLUSION}
	In summary, we studied the evolution of the anomalous Nernst and Hall conductivities in chemically doped Co$_3$Sn$_2$S$_2$ and found that  the amplitude of the anomalous transverse electric and thermoelectric conductivities evolve with doping. The low-temperature anomalous Hall conductivity varies much more than what is expected according to the calculated Berry spectrum. This indicates that the extrinsic contributions matter. The anomalous Nernst effect displays a more complex evolution as a function of doping and temperature. Nevertheless, the amplitude and the temperature dependence of $\alpha^A_{ij}/\sigma^A_{ij}$ does not vary much with doping and remains close to $\approx $ 0.5 $k_B/e$ near the Curie temperature.

	\section{ACKNOWLEDGMENTS}
	This work was supported by the National Science Foundation of China (Grant No.12004123, 51861135104 and No.11574097), the Fundamental Research Funds for the Central Universities (Grant no. 2019kfyXMBZ071) and the National Key Research and Development Program of China (Grant No.2022YFA1403503),  K. B was supported by the Agence Nationale de la Recherche (ANR-19-CE30-0014-04) and the Weizmann-CNRS collaboration program. X. L. acknowledges the China National Postdoctoral Program for Innovative Talents (Grant No.BX20200143) and the China Postdoctoral Science Foundation (Grant No.2020M682386).

	\noindent
	* \verb|zengwei.zhu@hust.edu.cn|\\

%

	
	\clearpage

	\renewcommand{\thesection}{S\arabic{section}}
	\renewcommand{\thetable}{S\arabic{table}}
	\renewcommand{\thefigure}{S\arabic{figure}}
	\renewcommand{\theequation}{S\arabic{equation}}
	
	\setcounter{section}{0}
	\setcounter{figure}{0}
	\setcounter{table}{0}
	\setcounter{equation}{0}
	

	
	

	

\begin{thebibliography}{42}%
	\makeatletter
	\providecommand \@ifxundefined [1]{%
		\@ifx{#1\undefined}
	}%
	\providecommand \@ifnum [1]{%
		\ifnum #1\expandafter \@firstoftwo
		\else \expandafter \@secondoftwo
		\fi
	}%
	\providecommand \@ifx [1]{%
		\ifx #1\expandafter \@firstoftwo
		\else \expandafter \@secondoftwo
		\fi
	}%
	\providecommand \natexlab [1]{#1}%
	\providecommand \enquote  [1]{``#1''}%
	\providecommand \bibnamefont  [1]{#1}%
	\providecommand \bibfnamefont [1]{#1}%
	\providecommand \citenamefont [1]{#1}%
	\providecommand \href@noop [0]{\@secondoftwo}%
	\providecommand \href [0]{\begingroup \@sanitize@url \@href}%
	\providecommand \@href[1]{\@@startlink{#1}\@@href}%
	\providecommand \@@href[1]{\endgroup#1\@@endlink}%
	\providecommand \@sanitize@url [0]{\catcode `\\12\catcode `\$12\catcode
		`\&12\catcode `\#12\catcode `\^12\catcode `\_12\catcode `\%12\relax}%
	\providecommand \@@startlink[1]{}%
	\providecommand \@@endlink[0]{}%
	\providecommand \url  [0]{\begingroup\@sanitize@url \@url }%
	\providecommand \@url [1]{\endgroup\@href {#1}{\urlprefix }}%
	\providecommand \urlprefix  [0]{URL }%
	\providecommand \Eprint [0]{\href }%
	\providecommand \doibase [0]{https://doi.org/}%
	\providecommand \selectlanguage [0]{\@gobble}%
	\providecommand \bibinfo  [0]{\@secondoftwo}%
	\providecommand \bibfield  [0]{\@secondoftwo}%
	\providecommand \translation [1]{[#1]}%
	\providecommand \BibitemOpen [0]{}%
	\providecommand \bibitemStop [0]{}%
	\providecommand \bibitemNoStop [0]{.\EOS\space}%
	\providecommand \EOS [0]{\spacefactor3000\relax}%
	\providecommand \BibitemShut  [1]{\csname bibitem#1\endcsname}%
	\let\auto@bib@innerbib\@empty
	\bibitem [{\citenamefont {Nagaosa}\ \emph {et~al.}(2010)\citenamefont
		{Nagaosa}, \citenamefont {Sinova}, \citenamefont {Onoda}, \citenamefont
		{MacDonald},\ and\ \citenamefont {Ong}}]{Nagaosa2010}%
	\BibitemOpen
	\bibfield  {author} {\bibinfo {author} {\bibfnamefont {N.}~\bibnamefont
			{Nagaosa}}, \bibinfo {author} {\bibfnamefont {J.}~\bibnamefont {Sinova}},
		\bibinfo {author} {\bibfnamefont {S.}~\bibnamefont {Onoda}}, \bibinfo
		{author} {\bibfnamefont {A.~H.}\ \bibnamefont {MacDonald}},\ and\ \bibinfo
		{author} {\bibfnamefont {N.~P.}\ \bibnamefont {Ong}},\ }\bibfield  {title}
	{\bibinfo {title} {Anomalous {Hall} effect},\ }\href
	{https://doi.org/10.1103/RevModPhys.82.1539} {\bibfield  {journal} {\bibinfo
			{journal} {Reviews of Modern Physics}\ }\textbf {\bibinfo {volume} {82}},\
		\bibinfo {pages} {1539} (\bibinfo {year} {2010})}\BibitemShut {NoStop}%
	\bibitem [{\citenamefont {Xiao}\ \emph {et~al.}(2010)\citenamefont {Xiao},
		\citenamefont {Chang},\ and\ \citenamefont {Niu}}]{Xiao2010}%
	\BibitemOpen
	\bibfield  {author} {\bibinfo {author} {\bibfnamefont {D.}~\bibnamefont
			{Xiao}}, \bibinfo {author} {\bibfnamefont {M.-C.}\ \bibnamefont {Chang}},\
		and\ \bibinfo {author} {\bibfnamefont {Q.}~\bibnamefont {Niu}},\ }\bibfield
	{title} {\bibinfo {title} {Berry phase effects on electronic properties},\
	}\href {https://doi.org/10.1103/RevModPhys.82.1959} {\bibfield  {journal}
		{\bibinfo  {journal} {Reviews of Modern Physics}\ }\textbf {\bibinfo {volume}
			{82}},\ \bibinfo {pages} {1959} (\bibinfo {year} {2010})}\BibitemShut
	{NoStop}%
	\bibitem [{\citenamefont {Xiao}\ \emph {et~al.}(2006)\citenamefont {Xiao},
		\citenamefont {Yao}, \citenamefont {Fang},\ and\ \citenamefont
		{Niu}}]{Xiao2006}%
	\BibitemOpen
	\bibfield  {author} {\bibinfo {author} {\bibfnamefont {D.}~\bibnamefont
			{Xiao}}, \bibinfo {author} {\bibfnamefont {Y.}~\bibnamefont {Yao}}, \bibinfo
		{author} {\bibfnamefont {Z.}~\bibnamefont {Fang}},\ and\ \bibinfo {author}
		{\bibfnamefont {Q.}~\bibnamefont {Niu}},\ }\bibfield  {title} {\bibinfo
		{title} {Berry-phase effect in anomalous thermoelectric transport},\ }\href
	{https://doi.org/10.1103/PhysRevLett.97.026603} {\bibfield  {journal}
		{\bibinfo  {journal} {Physical Review Letters}\ }\textbf {\bibinfo {volume}
			{97}},\ \bibinfo {pages} {026603} (\bibinfo {year} {2006})}\BibitemShut
	{NoStop}%
	\bibitem [{\citenamefont {Onoda}\ \emph {et~al.}(2008)\citenamefont {Onoda},
		\citenamefont {Sugimoto},\ and\ \citenamefont {Nagaosa}}]{Onoda2008}%
	\BibitemOpen
	\bibfield  {author} {\bibinfo {author} {\bibfnamefont {S.}~\bibnamefont
			{Onoda}}, \bibinfo {author} {\bibfnamefont {N.}~\bibnamefont {Sugimoto}},\
		and\ \bibinfo {author} {\bibfnamefont {N.}~\bibnamefont {Nagaosa}},\
	}\bibfield  {title} {\bibinfo {title} {Quantum transport theory of anomalous
			electric, thermoelectric, and thermal {Hall} effects in ferromagnets},\
	}\href {https://doi.org/10.1103/PhysRevB.77.165103} {\bibfield  {journal}
		{\bibinfo  {journal} {Phys. Rev. B}\ }\textbf {\bibinfo {volume} {77}},\
		\bibinfo {pages} {165103} (\bibinfo {year} {2008})}\BibitemShut {NoStop}%
	\bibitem [{\citenamefont {Pu}\ \emph {et~al.}(2008)\citenamefont {Pu},
		\citenamefont {Chiba}, \citenamefont {Matsukura}, \citenamefont {Ohno},\ and\
		\citenamefont {Shi}}]{Pu2008}%
	\BibitemOpen
	\bibfield  {author} {\bibinfo {author} {\bibfnamefont {Y.}~\bibnamefont
			{Pu}}, \bibinfo {author} {\bibfnamefont {D.}~\bibnamefont {Chiba}}, \bibinfo
		{author} {\bibfnamefont {F.}~\bibnamefont {Matsukura}}, \bibinfo {author}
		{\bibfnamefont {H.}~\bibnamefont {Ohno}},\ and\ \bibinfo {author}
		{\bibfnamefont {J.}~\bibnamefont {Shi}},\ }\bibfield  {title} {\bibinfo
		{title} {Mott relation for anomalous {Hall and Nernst} effects in
			${\mathrm{ga}}_{1\ensuremath{-}x}{\mathrm{mn}}_{x}\mathrm{As}$ ferromagnetic
			semiconductors},\ }\href {https://doi.org/10.1103/PhysRevLett.101.117208}
	{\bibfield  {journal} {\bibinfo  {journal} {Physical Review Letters}\
		}\textbf {\bibinfo {volume} {101}},\ \bibinfo {pages} {117208} (\bibinfo
		{year} {2008})}\BibitemShut {NoStop}%
	\bibitem [{\citenamefont {Li}\ \emph {et~al.}(2017)\citenamefont {Li},
		\citenamefont {Xu}, \citenamefont {Ding}, \citenamefont {Wang}, \citenamefont
		{Shen}, \citenamefont {Lu}, \citenamefont {Zhu},\ and\ \citenamefont
		{Behnia}}]{Li2017PRL}%
	\BibitemOpen
	\bibfield  {author} {\bibinfo {author} {\bibfnamefont {X.}~\bibnamefont
			{Li}}, \bibinfo {author} {\bibfnamefont {L.}~\bibnamefont {Xu}}, \bibinfo
		{author} {\bibfnamefont {L.}~\bibnamefont {Ding}}, \bibinfo {author}
		{\bibfnamefont {J.}~\bibnamefont {Wang}}, \bibinfo {author} {\bibfnamefont
			{M.}~\bibnamefont {Shen}}, \bibinfo {author} {\bibfnamefont {X.}~\bibnamefont
			{Lu}}, \bibinfo {author} {\bibfnamefont {Z.}~\bibnamefont {Zhu}},\ and\
		\bibinfo {author} {\bibfnamefont {K.}~\bibnamefont {Behnia}},\ }\bibfield
	{title} {\bibinfo {title} {Anomalous {Nernst and Righi-Leduc} effects in
			${\mathrm{mn}}_{3}\mathrm{Sn}$: Berry curvature and entropy flow},\ }\href
	{https://doi.org/10.1103/PhysRevLett.119.056601} {\bibfield  {journal}
		{\bibinfo  {journal} {Physical Review Letters}\ }\textbf {\bibinfo {volume}
			{119}},\ \bibinfo {pages} {056601} (\bibinfo {year} {2017})}\BibitemShut
	{NoStop}%
	\bibitem [{\citenamefont {Mott}\ and\ \citenamefont {Davis}(1971)}]{Mott1971}%
	\BibitemOpen
	\bibfield  {author} {\bibinfo {author} {\bibfnamefont {N.~F.}\ \bibnamefont
			{Mott}}\ and\ \bibinfo {author} {\bibfnamefont {E.~A.}\ \bibnamefont
			{Davis}},\ }\href@noop {} {\emph {\bibinfo {title} {Electronic Processes in
				Non-Crystalline Materials}}}\ (\bibinfo  {publisher} {Clarendon},\ \bibinfo
	{address} {Oxford},\ \bibinfo {year} {1971})\BibitemShut {NoStop}%
	\bibitem [{\citenamefont {Behnia}(2015)}]{Behnia2015b}%
	\BibitemOpen
	\bibfield  {author} {\bibinfo {author} {\bibfnamefont {K.}~\bibnamefont
			{Behnia}},\ }\href@noop {} {\emph {\bibinfo {title} {{Fundamentals of
					Thermoelectricity}}}}\ (\bibinfo  {publisher} {Oxford University Press},\
	\bibinfo {year} {2015})\BibitemShut {NoStop}%
	\bibitem [{\citenamefont {Behnia}\ and\ \citenamefont
		{Aubin}(2016)}]{Behnia2016}%
	\BibitemOpen
	\bibfield  {author} {\bibinfo {author} {\bibfnamefont {K.}~\bibnamefont
			{Behnia}}\ and\ \bibinfo {author} {\bibfnamefont {H.}~\bibnamefont {Aubin}},\
	}\bibfield  {title} {\bibinfo {title} {Nernst effect in metals and
			superconductors: a review of concepts and experiments},\ }\href
	{https://doi.org/10.1088/0034-4885/79/4/046502} {\bibfield  {journal}
		{\bibinfo  {journal} {Rep. Prog. Phys.}\ }\textbf {\bibinfo {volume} {79}},\
		\bibinfo {pages} {046502} (\bibinfo {year} {2016})}\BibitemShut {NoStop}%
	\bibitem [{\citenamefont {Xu}\ \emph {et~al.}(2020)\citenamefont {Xu},
		\citenamefont {Li}, \citenamefont {Ding}, \citenamefont {Chen}, \citenamefont
		{Sakai}, \citenamefont {Fauqué}, \citenamefont {Nakatsuji}, \citenamefont
		{Zhu},\ and\ \citenamefont {Behnia}}]{Xu2020}%
	\BibitemOpen
	\bibfield  {author} {\bibinfo {author} {\bibfnamefont {L.}~\bibnamefont
			{Xu}}, \bibinfo {author} {\bibfnamefont {X.}~\bibnamefont {Li}}, \bibinfo
		{author} {\bibfnamefont {L.}~\bibnamefont {Ding}}, \bibinfo {author}
		{\bibfnamefont {T.}~\bibnamefont {Chen}}, \bibinfo {author} {\bibfnamefont
			{A.}~\bibnamefont {Sakai}}, \bibinfo {author} {\bibfnamefont
			{B.}~\bibnamefont {Fauqué}}, \bibinfo {author} {\bibfnamefont
			{S.}~\bibnamefont {Nakatsuji}}, \bibinfo {author} {\bibfnamefont
			{Z.}~\bibnamefont {Zhu}},\ and\ \bibinfo {author} {\bibfnamefont
			{K.}~\bibnamefont {Behnia}},\ }\bibfield  {title} {\bibinfo {title}
		{Anomalous transverse response of ${\mathrm{co}}_{2}\mathrm{MnGa}$ and
			universality of the room-temperature
			${\ensuremath{\alpha}}_{ij}^{A}/{\ensuremath{\sigma}}_{ij}^{A}$ ratio across
			topological magnets},\ }\href {https://doi.org/10.1103/PhysRevB.101.180404}
	{\bibfield  {journal} {\bibinfo  {journal} {Physical Review B}\ }\textbf
		{\bibinfo {volume} {101}},\ \bibinfo {pages} {180404} (\bibinfo {year}
		{2020})}\BibitemShut {NoStop}%
	\bibitem [{\citenamefont {Guo}\ \emph {et~al.}(2017)\citenamefont {Guo},
		\citenamefont {Ou}, \citenamefont {Xu}, \citenamefont {Feng}, \citenamefont
		{Jiang}, \citenamefont {He}, \citenamefont {Ma}, \citenamefont {Xue},\ and\
		\citenamefont {Wang}}]{Guo2017}%
	\BibitemOpen
	\bibfield  {author} {\bibinfo {author} {\bibfnamefont {M.}~\bibnamefont
			{Guo}}, \bibinfo {author} {\bibfnamefont {Y.}~\bibnamefont {Ou}}, \bibinfo
		{author} {\bibfnamefont {Y.}~\bibnamefont {Xu}}, \bibinfo {author}
		{\bibfnamefont {Y.}~\bibnamefont {Feng}}, \bibinfo {author} {\bibfnamefont
			{G.}~\bibnamefont {Jiang}}, \bibinfo {author} {\bibfnamefont
			{K.}~\bibnamefont {He}}, \bibinfo {author} {\bibfnamefont {X.}~\bibnamefont
			{Ma}}, \bibinfo {author} {\bibfnamefont {Q.-K.}\ \bibnamefont {Xue}},\ and\
		\bibinfo {author} {\bibfnamefont {Y.}~\bibnamefont {Wang}},\ }\bibfield
	{title} {\bibinfo {title} {Ambi-polar anomalous {Nernst} effect in a magnetic
			topological insulator},\ }\href {https://doi.org/10.1088/1367-2630/aa8b91}
	{\bibfield  {journal} {\bibinfo  {journal} {New Journal of Physics}\ }\textbf
		{\bibinfo {volume} {19}},\ \bibinfo {pages} {113009} (\bibinfo {year}
		{2017})}\BibitemShut {NoStop}%
	\bibitem [{\citenamefont {Miyasato}\ \emph {et~al.}(2007)\citenamefont
		{Miyasato}, \citenamefont {Abe}, \citenamefont {Fujii}, \citenamefont
		{Asamitsu}, \citenamefont {Onoda}, \citenamefont {Onose}, \citenamefont
		{Nagaosa},\ and\ \citenamefont {Tokura}}]{Miyasato2007}%
	\BibitemOpen
	\bibfield  {author} {\bibinfo {author} {\bibfnamefont {T.}~\bibnamefont
			{Miyasato}}, \bibinfo {author} {\bibfnamefont {N.}~\bibnamefont {Abe}},
		\bibinfo {author} {\bibfnamefont {T.}~\bibnamefont {Fujii}}, \bibinfo
		{author} {\bibfnamefont {A.}~\bibnamefont {Asamitsu}}, \bibinfo {author}
		{\bibfnamefont {S.}~\bibnamefont {Onoda}}, \bibinfo {author} {\bibfnamefont
			{Y.}~\bibnamefont {Onose}}, \bibinfo {author} {\bibfnamefont
			{N.}~\bibnamefont {Nagaosa}},\ and\ \bibinfo {author} {\bibfnamefont
			{Y.}~\bibnamefont {Tokura}},\ }\bibfield  {title} {\bibinfo {title}
		{Crossover behavior of the anomalous {Hall} effect and anomalous {Nernst}
			effect in itinerant ferromagnets},\ }\href
	{https://doi.org/10.1103/PhysRevLett.99.086602} {\bibfield  {journal}
		{\bibinfo  {journal} {Physical Review Letters}\ }\textbf {\bibinfo {volume}
			{99}},\ \bibinfo {pages} {086602} (\bibinfo {year} {2007})}\BibitemShut
	{NoStop}%
	\bibitem [{\citenamefont {Breidenbach}\ \emph {et~al.}(2022)\citenamefont
		{Breidenbach}, \citenamefont {Yu}, \citenamefont {Peterson}, \citenamefont
		{McFadden}, \citenamefont {Peria}, \citenamefont {Palmstr\o{}m},\ and\
		\citenamefont {Crowell}}]{Breidenbach2022}%
	\BibitemOpen
	\bibfield  {author} {\bibinfo {author} {\bibfnamefont {A.~T.}\ \bibnamefont
			{Breidenbach}}, \bibinfo {author} {\bibfnamefont {H.}~\bibnamefont {Yu}},
		\bibinfo {author} {\bibfnamefont {T.~A.}\ \bibnamefont {Peterson}}, \bibinfo
		{author} {\bibfnamefont {A.~P.}\ \bibnamefont {McFadden}}, \bibinfo {author}
		{\bibfnamefont {W.~K.}\ \bibnamefont {Peria}}, \bibinfo {author}
		{\bibfnamefont {C.~J.}\ \bibnamefont {Palmstr\o{}m}},\ and\ \bibinfo {author}
		{\bibfnamefont {P.~A.}\ \bibnamefont {Crowell}},\ }\bibfield  {title}
	{\bibinfo {title} {Anomalous {Nernst and Seebeck} coefficients in epitaxial
			thin film
			{${\mathrm{Co}}_{2}{\mathrm{MnAl}}_{x}{\mathrm{Si}}_{1\ensuremath{-}x}$ and
				${\mathrm{Co}}_{2}\mathrm{FeAl}$}},\ }\href
	{https://doi.org/10.1103/PhysRevB.105.144405} {\bibfield  {journal} {\bibinfo
			{journal} {Phys. Rev. B}\ }\textbf {\bibinfo {volume} {105}},\ \bibinfo
		{pages} {144405} (\bibinfo {year} {2022})}\BibitemShut {NoStop}%
	\bibitem [{\citenamefont {Liu}\ \emph {et~al.}(2018)\citenamefont {Liu},
		\citenamefont {Sun}, \citenamefont {Kumar}, \citenamefont {Muechler},
		\citenamefont {Sun}, \citenamefont {Jiao}, \citenamefont {Yang},
		\citenamefont {Liu}, \citenamefont {Liang}, \citenamefont {Xu}, \citenamefont
		{Kroder}, \citenamefont {Süß}, \citenamefont {Borrmann}, \citenamefont
		{Shekhar}, \citenamefont {Wang}, \citenamefont {Xi}, \citenamefont {Wang},
		\citenamefont {Schnelle}, \citenamefont {Wirth}, \citenamefont {Chen},
		\citenamefont {Goennenwein},\ and\ \citenamefont {Felser}}]{Liu2018CSS}%
	\BibitemOpen
	\bibfield  {author} {\bibinfo {author} {\bibfnamefont {E.}~\bibnamefont
			{Liu}}, \bibinfo {author} {\bibfnamefont {Y.}~\bibnamefont {Sun}}, \bibinfo
		{author} {\bibfnamefont {N.}~\bibnamefont {Kumar}}, \bibinfo {author}
		{\bibfnamefont {L.}~\bibnamefont {Muechler}}, \bibinfo {author}
		{\bibfnamefont {A.}~\bibnamefont {Sun}}, \bibinfo {author} {\bibfnamefont
			{L.}~\bibnamefont {Jiao}}, \bibinfo {author} {\bibfnamefont {S.-Y.}\
			\bibnamefont {Yang}}, \bibinfo {author} {\bibfnamefont {D.}~\bibnamefont
			{Liu}}, \bibinfo {author} {\bibfnamefont {A.}~\bibnamefont {Liang}}, \bibinfo
		{author} {\bibfnamefont {Q.}~\bibnamefont {Xu}}, \bibinfo {author}
		{\bibfnamefont {J.}~\bibnamefont {Kroder}}, \bibinfo {author} {\bibfnamefont
			{V.}~\bibnamefont {Süß}}, \bibinfo {author} {\bibfnamefont
			{H.}~\bibnamefont {Borrmann}}, \bibinfo {author} {\bibfnamefont
			{C.}~\bibnamefont {Shekhar}}, \bibinfo {author} {\bibfnamefont
			{Z.}~\bibnamefont {Wang}}, \bibinfo {author} {\bibfnamefont {C.}~\bibnamefont
			{Xi}}, \bibinfo {author} {\bibfnamefont {W.}~\bibnamefont {Wang}}, \bibinfo
		{author} {\bibfnamefont {W.}~\bibnamefont {Schnelle}}, \bibinfo {author}
		{\bibfnamefont {S.}~\bibnamefont {Wirth}}, \bibinfo {author} {\bibfnamefont
			{Y.}~\bibnamefont {Chen}}, \bibinfo {author} {\bibfnamefont {S.~T.~B.}\
			\bibnamefont {Goennenwein}},\ and\ \bibinfo {author} {\bibfnamefont
			{C.}~\bibnamefont {Felser}},\ }\bibfield  {title} {\bibinfo {title} {Giant
			anomalous {Hall} effect in a ferromagnetic kagome-lattice semimetal},\ }\href
	{https://doi.org/10.1038/s41567-018-0234-5} {\bibfield  {journal} {\bibinfo
			{journal} {Nature Physics}\ }\textbf {\bibinfo {volume} {14}},\ \bibinfo
		{pages} {1125} (\bibinfo {year} {2018})}\BibitemShut {NoStop}%
	\bibitem [{\citenamefont {Wang}\ \emph {et~al.}(2018)\citenamefont {Wang},
		\citenamefont {Xu}, \citenamefont {Lou}, \citenamefont {Liu}, \citenamefont
		{Li}, \citenamefont {Huang}, \citenamefont {Shen}, \citenamefont {Weng},
		\citenamefont {Wang},\ and\ \citenamefont {Lei}}]{Wang2018}%
	\BibitemOpen
	\bibfield  {author} {\bibinfo {author} {\bibfnamefont {Q.}~\bibnamefont
			{Wang}}, \bibinfo {author} {\bibfnamefont {Y.}~\bibnamefont {Xu}}, \bibinfo
		{author} {\bibfnamefont {R.}~\bibnamefont {Lou}}, \bibinfo {author}
		{\bibfnamefont {Z.}~\bibnamefont {Liu}}, \bibinfo {author} {\bibfnamefont
			{M.}~\bibnamefont {Li}}, \bibinfo {author} {\bibfnamefont {Y.}~\bibnamefont
			{Huang}}, \bibinfo {author} {\bibfnamefont {D.}~\bibnamefont {Shen}},
		\bibinfo {author} {\bibfnamefont {H.}~\bibnamefont {Weng}}, \bibinfo {author}
		{\bibfnamefont {S.}~\bibnamefont {Wang}},\ and\ \bibinfo {author}
		{\bibfnamefont {H.}~\bibnamefont {Lei}},\ }\bibfield  {title} {\bibinfo
		{title} {Large intrinsic anomalous {Hall} effect in half-metallic ferromagnet
			{Co$_3$Sn$_2$S$_2$} with magnetic {Weyl} fermions},\ }\href
	{https://doi.org/10.1038/s41467-018-06088-2} {\bibfield  {journal} {\bibinfo
			{journal} {Nature Communications}\ }\textbf {\bibinfo {volume} {9}},\
		\bibinfo {pages} {3681} (\bibinfo {year} {2018})}\BibitemShut {NoStop}%
	\bibitem [{\citenamefont {Shen}\ \emph
		{et~al.}(2020{\natexlab{a}})\citenamefont {Shen}, \citenamefont {Yao},
		\citenamefont {Zeng}, \citenamefont {Sun}, \citenamefont {Xi}, \citenamefont
		{Wu}, \citenamefont {Wang}, \citenamefont {Shen}, \citenamefont {Liu},\ and\
		\citenamefont {Liu}}]{Shen2020PRL}%
	\BibitemOpen
	\bibfield  {author} {\bibinfo {author} {\bibfnamefont {J.}~\bibnamefont
			{Shen}}, \bibinfo {author} {\bibfnamefont {Q.}~\bibnamefont {Yao}}, \bibinfo
		{author} {\bibfnamefont {Q.}~\bibnamefont {Zeng}}, \bibinfo {author}
		{\bibfnamefont {H.}~\bibnamefont {Sun}}, \bibinfo {author} {\bibfnamefont
			{X.}~\bibnamefont {Xi}}, \bibinfo {author} {\bibfnamefont {G.}~\bibnamefont
			{Wu}}, \bibinfo {author} {\bibfnamefont {W.}~\bibnamefont {Wang}}, \bibinfo
		{author} {\bibfnamefont {B.}~\bibnamefont {Shen}}, \bibinfo {author}
		{\bibfnamefont {Q.}~\bibnamefont {Liu}},\ and\ \bibinfo {author}
		{\bibfnamefont {E.}~\bibnamefont {Liu}},\ }\bibfield  {title} {\bibinfo
		{title} {Local disorder-induced elevation of intrinsic anomalous {Hall}
			conductance in an electron-doped magnetic {Weyl} semimetal},\ }\href
	{https://doi.org/10.1103/PhysRevLett.125.086602} {\bibfield  {journal}
		{\bibinfo  {journal} {Physical Review Letters}\ }\textbf {\bibinfo {volume}
			{125}},\ \bibinfo {pages} {086602} (\bibinfo {year}
		{2020}{\natexlab{a}})}\BibitemShut {NoStop}%
	\bibitem [{\citenamefont {Muechler}\ \emph {et~al.}(2020)\citenamefont
		{Muechler}, \citenamefont {Liu}, \citenamefont {Gayles}, \citenamefont {Xu},
		\citenamefont {Felser},\ and\ \citenamefont {Sun}}]{Muechler2020}%
	\BibitemOpen
	\bibfield  {author} {\bibinfo {author} {\bibfnamefont {L.}~\bibnamefont
			{Muechler}}, \bibinfo {author} {\bibfnamefont {E.}~\bibnamefont {Liu}},
		\bibinfo {author} {\bibfnamefont {J.}~\bibnamefont {Gayles}}, \bibinfo
		{author} {\bibfnamefont {Q.}~\bibnamefont {Xu}}, \bibinfo {author}
		{\bibfnamefont {C.}~\bibnamefont {Felser}},\ and\ \bibinfo {author}
		{\bibfnamefont {Y.}~\bibnamefont {Sun}},\ }\bibfield  {title} {\bibinfo
		{title} {Emerging chiral edge states from the confinement of a magnetic
			{Weyl} semimetal in {${\mathrm{Co}}_{3}{\mathrm{Sn}}_{2}{\mathrm{S}}_{2}$}},\
	}\href {https://doi.org/10.1103/PhysRevB.101.115106} {\bibfield  {journal}
		{\bibinfo  {journal} {Physical Review B}\ }\textbf {\bibinfo {volume}
			{101}},\ \bibinfo {pages} {115106} (\bibinfo {year} {2020})}\BibitemShut
	{NoStop}%
	\bibitem [{\citenamefont {Shen}\ \emph
		{et~al.}(2020{\natexlab{b}})\citenamefont {Shen}, \citenamefont {Zeng},
		\citenamefont {Zhang}, \citenamefont {Sun}, \citenamefont {Yao},
		\citenamefont {Xi}, \citenamefont {Wang}, \citenamefont {Wu}, \citenamefont
		{Shen}, \citenamefont {Liu},\ and\ \citenamefont {Liu}}]{Shen2020}%
	\BibitemOpen
	\bibfield  {author} {\bibinfo {author} {\bibfnamefont {J.}~\bibnamefont
			{Shen}}, \bibinfo {author} {\bibfnamefont {Q.}~\bibnamefont {Zeng}}, \bibinfo
		{author} {\bibfnamefont {S.}~\bibnamefont {Zhang}}, \bibinfo {author}
		{\bibfnamefont {H.}~\bibnamefont {Sun}}, \bibinfo {author} {\bibfnamefont
			{Q.}~\bibnamefont {Yao}}, \bibinfo {author} {\bibfnamefont {X.}~\bibnamefont
			{Xi}}, \bibinfo {author} {\bibfnamefont {W.}~\bibnamefont {Wang}}, \bibinfo
		{author} {\bibfnamefont {G.}~\bibnamefont {Wu}}, \bibinfo {author}
		{\bibfnamefont {B.}~\bibnamefont {Shen}}, \bibinfo {author} {\bibfnamefont
			{Q.}~\bibnamefont {Liu}},\ and\ \bibinfo {author} {\bibfnamefont
			{E.}~\bibnamefont {Liu}},\ }\bibfield  {title} {\bibinfo {title} {33\% giant
			anomalous {Hall} current driven by both intrinsic and extrinsic contributions
			in magnetic {Weyl} semimetal {Co$_3$Sn$_2$S$_2$}},\ }\href
	{https://doi.org/https://doi.org/10.1002/adfm.202000830} {\bibfield
		{journal} {\bibinfo  {journal} {Advanced Functional Materials}\ }\textbf
		{\bibinfo {volume} {30}},\ \bibinfo {pages} {2000830} (\bibinfo {year}
		{2020}{\natexlab{b}})}\BibitemShut {NoStop}%
	\bibitem [{\citenamefont {Corps}\ \emph {et~al.}(2015)\citenamefont {Corps},
		\citenamefont {Vaqueiro}, \citenamefont {Aziz}, \citenamefont {Grau-Crespo},
		\citenamefont {Kockelmann}, \citenamefont {Jumas},\ and\ \citenamefont
		{Powell}}]{Corps2015}%
	\BibitemOpen
	\bibfield  {author} {\bibinfo {author} {\bibfnamefont {J.}~\bibnamefont
			{Corps}}, \bibinfo {author} {\bibfnamefont {P.}~\bibnamefont {Vaqueiro}},
		\bibinfo {author} {\bibfnamefont {A.}~\bibnamefont {Aziz}}, \bibinfo {author}
		{\bibfnamefont {R.}~\bibnamefont {Grau-Crespo}}, \bibinfo {author}
		{\bibfnamefont {W.}~\bibnamefont {Kockelmann}}, \bibinfo {author}
		{\bibfnamefont {J.-C.}\ \bibnamefont {Jumas}},\ and\ \bibinfo {author}
		{\bibfnamefont {A.~V.}\ \bibnamefont {Powell}},\ }\bibfield  {title}
	{\bibinfo {title} {Interplay of metal-atom ordering, fermi level tuning, and
			thermoelectric properties in cobalt shandites {Co$_3$M$_2$S$_2$ (M = Sn,
				In)}},\ }\href {https://doi.org/10.1021/acs.chemmater.5b00801} {\bibfield
		{journal} {\bibinfo  {journal} {Chemistry of Materials}\ }\textbf {\bibinfo
			{volume} {27}},\ \bibinfo {pages} {3946} (\bibinfo {year}
		{2015})}\BibitemShut {NoStop}%
	\bibitem [{\citenamefont {Zhou}\ \emph {et~al.}(2020)\citenamefont {Zhou},
		\citenamefont {Chang}, \citenamefont {Wang}, \citenamefont {Gui},
		\citenamefont {Xu}, \citenamefont {Yin}, \citenamefont {Guguchia},
		\citenamefont {Zhang}, \citenamefont {Chang}, \citenamefont {Lin},
		\citenamefont {Xie}, \citenamefont {Hasan},\ and\ \citenamefont
		{Jia}}]{Zhou2020PRB}%
	\BibitemOpen
	\bibfield  {author} {\bibinfo {author} {\bibfnamefont {H.}~\bibnamefont
			{Zhou}}, \bibinfo {author} {\bibfnamefont {G.}~\bibnamefont {Chang}},
		\bibinfo {author} {\bibfnamefont {G.}~\bibnamefont {Wang}}, \bibinfo {author}
		{\bibfnamefont {X.}~\bibnamefont {Gui}}, \bibinfo {author} {\bibfnamefont
			{X.}~\bibnamefont {Xu}}, \bibinfo {author} {\bibfnamefont {J.-X.}\
			\bibnamefont {Yin}}, \bibinfo {author} {\bibfnamefont {Z.}~\bibnamefont
			{Guguchia}}, \bibinfo {author} {\bibfnamefont {S.~S.}\ \bibnamefont {Zhang}},
		\bibinfo {author} {\bibfnamefont {T.-R.}\ \bibnamefont {Chang}}, \bibinfo
		{author} {\bibfnamefont {H.}~\bibnamefont {Lin}}, \bibinfo {author}
		{\bibfnamefont {W.}~\bibnamefont {Xie}}, \bibinfo {author} {\bibfnamefont
			{M.~Z.}\ \bibnamefont {Hasan}},\ and\ \bibinfo {author} {\bibfnamefont
			{S.}~\bibnamefont {Jia}},\ }\bibfield  {title} {\bibinfo {title} {Enhanced
			anomalous {Hall} effect in the magnetic topological semimetal
			{$\mathrm{Co}{}_{3}{\mathrm{Sn}}_{2\ensuremath{-}x}{\mathrm{In}}_{x}{\mathrm{S}}_{2}$}},\
	}\href {https://doi.org/10.1103/PhysRevB.101.125121} {\bibfield  {journal}
		{\bibinfo  {journal} {Physical Review B}\ }\textbf {\bibinfo {volume}
			{101}},\ \bibinfo {pages} {125121} (\bibinfo {year} {2020})}\BibitemShut
	{NoStop}%
	\bibitem [{\citenamefont {Thakur}\ \emph {et~al.}(2020)\citenamefont {Thakur},
		\citenamefont {Vir}, \citenamefont {Guin}, \citenamefont {Shekhar},
		\citenamefont {Weihrich}, \citenamefont {Sun}, \citenamefont {Kumar},\ and\
		\citenamefont {Felser}}]{Thakur2020}%
	\BibitemOpen
	\bibfield  {author} {\bibinfo {author} {\bibfnamefont {G.~S.}\ \bibnamefont
			{Thakur}}, \bibinfo {author} {\bibfnamefont {P.}~\bibnamefont {Vir}},
		\bibinfo {author} {\bibfnamefont {S.~N.}\ \bibnamefont {Guin}}, \bibinfo
		{author} {\bibfnamefont {C.}~\bibnamefont {Shekhar}}, \bibinfo {author}
		{\bibfnamefont {R.}~\bibnamefont {Weihrich}}, \bibinfo {author}
		{\bibfnamefont {Y.}~\bibnamefont {Sun}}, \bibinfo {author} {\bibfnamefont
			{N.}~\bibnamefont {Kumar}},\ and\ \bibinfo {author} {\bibfnamefont
			{C.}~\bibnamefont {Felser}},\ }\bibfield  {title} {\bibinfo {title}
		{Intrinsic anomalous {Hall} effect in {Ni}-substituted magnetic {Weyl}
			semimetal {Co$_3$Sn$_2$S$_2$}},\ }\href
	{https://doi.org/10.1021/acs.chemmater.9b05009} {\bibfield  {journal}
		{\bibinfo  {journal} {Chemistry of Materials}\ }\textbf {\bibinfo {volume}
			{32}},\ \bibinfo {pages} {1612} (\bibinfo {year} {2020})}\BibitemShut
	{NoStop}%
	\bibitem [{\citenamefont {Yanagi}\ \emph {et~al.}(2021)\citenamefont {Yanagi},
		\citenamefont {Ikeda}, \citenamefont {Fujiwara}, \citenamefont {Nomura},
		\citenamefont {Tsukazaki},\ and\ \citenamefont {Suzuki}}]{Yanagi2021}%
	\BibitemOpen
	\bibfield  {author} {\bibinfo {author} {\bibfnamefont {Y.}~\bibnamefont
			{Yanagi}}, \bibinfo {author} {\bibfnamefont {J.}~\bibnamefont {Ikeda}},
		\bibinfo {author} {\bibfnamefont {K.}~\bibnamefont {Fujiwara}}, \bibinfo
		{author} {\bibfnamefont {K.}~\bibnamefont {Nomura}}, \bibinfo {author}
		{\bibfnamefont {A.}~\bibnamefont {Tsukazaki}},\ and\ \bibinfo {author}
		{\bibfnamefont {M.-T.}\ \bibnamefont {Suzuki}},\ }\bibfield  {title}
	{\bibinfo {title} {First-principles investigation of magnetic and transport
			properties in hole-doped shandite compounds
			{${\mathrm{Co}}_{3}{\mathrm{In}}_{x}{\mathrm{Sn}}_{2\ensuremath{-}x}{\mathrm{S}}_{2}$}},\
	}\href {https://doi.org/10.1103/PhysRevB.103.205112} {\bibfield  {journal}
		{\bibinfo  {journal} {Physical Review B}\ }\textbf {\bibinfo {volume}
			{103}},\ \bibinfo {pages} {205112} (\bibinfo {year} {2021})}\BibitemShut
	{NoStop}%
	\bibitem [{\citenamefont {Chen}\ \emph {et~al.}(2019)\citenamefont {Chen},
		\citenamefont {Wang}, \citenamefont {Gu}, \citenamefont {Wang}, \citenamefont
		{Zhou}, \citenamefont {An}, \citenamefont {Zhou}, \citenamefont {Zhang},
		\citenamefont {Chen}, \citenamefont {Yuan}, \citenamefont {Qi}, \citenamefont
		{Zhang}, \citenamefont {Zhou}, \citenamefont {Zhou}, \citenamefont {Yao},\
		and\ \citenamefont {Yang}}]{Chen2019PRB}%
	\BibitemOpen
	\bibfield  {author} {\bibinfo {author} {\bibfnamefont {X.}~\bibnamefont
			{Chen}}, \bibinfo {author} {\bibfnamefont {M.}~\bibnamefont {Wang}}, \bibinfo
		{author} {\bibfnamefont {C.}~\bibnamefont {Gu}}, \bibinfo {author}
		{\bibfnamefont {S.}~\bibnamefont {Wang}}, \bibinfo {author} {\bibfnamefont
			{Y.}~\bibnamefont {Zhou}}, \bibinfo {author} {\bibfnamefont {C.}~\bibnamefont
			{An}}, \bibinfo {author} {\bibfnamefont {Y.}~\bibnamefont {Zhou}}, \bibinfo
		{author} {\bibfnamefont {B.}~\bibnamefont {Zhang}}, \bibinfo {author}
		{\bibfnamefont {C.}~\bibnamefont {Chen}}, \bibinfo {author} {\bibfnamefont
			{Y.}~\bibnamefont {Yuan}}, \bibinfo {author} {\bibfnamefont {M.}~\bibnamefont
			{Qi}}, \bibinfo {author} {\bibfnamefont {L.}~\bibnamefont {Zhang}}, \bibinfo
		{author} {\bibfnamefont {H.}~\bibnamefont {Zhou}}, \bibinfo {author}
		{\bibfnamefont {J.}~\bibnamefont {Zhou}}, \bibinfo {author} {\bibfnamefont
			{Y.}~\bibnamefont {Yao}},\ and\ \bibinfo {author} {\bibfnamefont
			{Z.}~\bibnamefont {Yang}},\ }\bibfield  {title} {\bibinfo {title}
		{Pressure-tunable large anomalous {Hall} effect of the ferromagnetic
			kagome-lattice {Weyl} semimetal
			{${\mathrm{Co}}_{3}{\mathrm{Sn}}_{2}{\mathrm{S}}_{2}$}},\ }\href
	{https://doi.org/10.1103/PhysRevB.100.165145} {\bibfield  {journal} {\bibinfo
			{journal} {Phys. Rev. B}\ }\textbf {\bibinfo {volume} {100}},\ \bibinfo
		{pages} {165145} (\bibinfo {year} {2019})}\BibitemShut {NoStop}%
	\bibitem [{\citenamefont {Liu}\ \emph {et~al.}(2020)\citenamefont {Liu},
		\citenamefont {Zhang}, \citenamefont {Xu}, \citenamefont {Yang},
		\citenamefont {Wang}, \citenamefont {Lei}, \citenamefont {Sui}, \citenamefont
		{Uwatoko}, \citenamefont {Wang}, \citenamefont {Weng}, \citenamefont {Sun},\
		and\ \citenamefont {Cheng}}]{Liu2020PRM}%
	\BibitemOpen
	\bibfield  {author} {\bibinfo {author} {\bibfnamefont {Z.~Y.}\ \bibnamefont
			{Liu}}, \bibinfo {author} {\bibfnamefont {T.}~\bibnamefont {Zhang}}, \bibinfo
		{author} {\bibfnamefont {S.~X.}\ \bibnamefont {Xu}}, \bibinfo {author}
		{\bibfnamefont {P.~T.}\ \bibnamefont {Yang}}, \bibinfo {author}
		{\bibfnamefont {Q.}~\bibnamefont {Wang}}, \bibinfo {author} {\bibfnamefont
			{H.~C.}\ \bibnamefont {Lei}}, \bibinfo {author} {\bibfnamefont
			{Y.}~\bibnamefont {Sui}}, \bibinfo {author} {\bibfnamefont {Y.}~\bibnamefont
			{Uwatoko}}, \bibinfo {author} {\bibfnamefont {B.~S.}\ \bibnamefont {Wang}},
		\bibinfo {author} {\bibfnamefont {H.~M.}\ \bibnamefont {Weng}}, \bibinfo
		{author} {\bibfnamefont {J.~P.}\ \bibnamefont {Sun}},\ and\ \bibinfo {author}
		{\bibfnamefont {J.-G.}\ \bibnamefont {Cheng}},\ }\bibfield  {title} {\bibinfo
		{title} {Pressure effect on the anomalous {Hall} effect of ferromagnetic
			{Weyl} semimetal
			{$\mathrm{C}{\mathrm{o}}_{3}\mathrm{S}{\mathrm{n}}_{2}{\mathrm{S}}_{2}$}},\
	}\href {https://doi.org/10.1103/PhysRevMaterials.4.044203} {\bibfield
		{journal} {\bibinfo  {journal} {Phys. Rev. Materials}\ }\textbf {\bibinfo
			{volume} {4}},\ \bibinfo {pages} {044203} (\bibinfo {year}
		{2020})}\BibitemShut {NoStop}%
	\bibitem [{\citenamefont {Zeng}\ \emph {et~al.}(2022)\citenamefont {Zeng},
		\citenamefont {Sun}, \citenamefont {Shen}, \citenamefont {Yao}, \citenamefont
		{Zhang}, \citenamefont {Li}, \citenamefont {Jiao}, \citenamefont {Wei},
		\citenamefont {Felser}, \citenamefont {Wang}, \citenamefont {Liu},\ and\
		\citenamefont {Liu}}]{Zeng2022}%
	\BibitemOpen
	\bibfield  {author} {\bibinfo {author} {\bibfnamefont {Q.}~\bibnamefont
			{Zeng}}, \bibinfo {author} {\bibfnamefont {H.}~\bibnamefont {Sun}}, \bibinfo
		{author} {\bibfnamefont {J.}~\bibnamefont {Shen}}, \bibinfo {author}
		{\bibfnamefont {Q.}~\bibnamefont {Yao}}, \bibinfo {author} {\bibfnamefont
			{Q.}~\bibnamefont {Zhang}}, \bibinfo {author} {\bibfnamefont
			{N.}~\bibnamefont {Li}}, \bibinfo {author} {\bibfnamefont {L.}~\bibnamefont
			{Jiao}}, \bibinfo {author} {\bibfnamefont {H.}~\bibnamefont {Wei}}, \bibinfo
		{author} {\bibfnamefont {C.}~\bibnamefont {Felser}}, \bibinfo {author}
		{\bibfnamefont {Y.}~\bibnamefont {Wang}}, \bibinfo {author} {\bibfnamefont
			{Q.}~\bibnamefont {Liu}},\ and\ \bibinfo {author} {\bibfnamefont
			{E.}~\bibnamefont {Liu}},\ }\bibfield  {title} {\bibinfo {title}
		{Pressure-driven magneto-topological phase transition in a magnetic {Weyl}
			semimetal},\ }\href {https://doi.org/https://doi.org/10.1002/qute.202100149}
	{\bibfield  {journal} {\bibinfo  {journal} {Advanced Quantum Technologies}\
		}\textbf {\bibinfo {volume} {5}},\ \bibinfo {pages} {2100149} (\bibinfo
		{year} {2022})}\BibitemShut {NoStop}%
	\bibitem [{\citenamefont {Guguchia}\ \emph {et~al.}(2021)\citenamefont
		{Guguchia}, \citenamefont {Zhou}, \citenamefont {Wang}, \citenamefont {Yin},
		\citenamefont {Mielke}, \citenamefont {Tsirkin}, \citenamefont {Belopolski},
		\citenamefont {Zhang}, \citenamefont {Cochran}, \citenamefont {Neupert},
		\citenamefont {Khasanov}, \citenamefont {Amato}, \citenamefont {Jia},
		\citenamefont {Hasan},\ and\ \citenamefont {Luetkens}}]{Guguchia2021}%
	\BibitemOpen
	\bibfield  {author} {\bibinfo {author} {\bibfnamefont {Z.}~\bibnamefont
			{Guguchia}}, \bibinfo {author} {\bibfnamefont {H.}~\bibnamefont {Zhou}},
		\bibinfo {author} {\bibfnamefont {C.~N.}\ \bibnamefont {Wang}}, \bibinfo
		{author} {\bibfnamefont {J.~X.}\ \bibnamefont {Yin}}, \bibinfo {author}
		{\bibfnamefont {C.}~\bibnamefont {Mielke}}, \bibinfo {author} {\bibfnamefont
			{S.~S.}\ \bibnamefont {Tsirkin}}, \bibinfo {author} {\bibfnamefont
			{I.}~\bibnamefont {Belopolski}}, \bibinfo {author} {\bibfnamefont {S.~S.}\
			\bibnamefont {Zhang}}, \bibinfo {author} {\bibfnamefont {T.~A.}\ \bibnamefont
			{Cochran}}, \bibinfo {author} {\bibfnamefont {T.}~\bibnamefont {Neupert}},
		\bibinfo {author} {\bibfnamefont {R.}~\bibnamefont {Khasanov}}, \bibinfo
		{author} {\bibfnamefont {A.}~\bibnamefont {Amato}}, \bibinfo {author}
		{\bibfnamefont {S.}~\bibnamefont {Jia}}, \bibinfo {author} {\bibfnamefont
			{M.~Z.}\ \bibnamefont {Hasan}},\ and\ \bibinfo {author} {\bibfnamefont
			{H.}~\bibnamefont {Luetkens}},\ }\bibfield  {title} {\bibinfo {title}
		{Multiple quantum phase transitions of different nature in the topological
			kagome magnet {Co$_3$Sn$_{2-x}$In$_x$S$_2$}},\ }\href
	{https://doi.org/10.1038/s41535-021-00352-3} {\bibfield  {journal} {\bibinfo
			{journal} {npj Quantum Materials}\ }\textbf {\bibinfo {volume} {6}},\
		\bibinfo {pages} {50} (\bibinfo {year} {2021})}\BibitemShut {NoStop}%
	\bibitem [{\citenamefont {Fujioka}\ \emph {et~al.}(2014)\citenamefont
		{Fujioka}, \citenamefont {Shibuya}, \citenamefont {Nakai}, \citenamefont
		{Yoshiyasu}, \citenamefont {Sakai}, \citenamefont {Takano}, \citenamefont
		{Kamihara},\ and\ \citenamefont {Matoba}}]{Fujioka2014}%
	\BibitemOpen
	\bibfield  {author} {\bibinfo {author} {\bibfnamefont {M.}~\bibnamefont
			{Fujioka}}, \bibinfo {author} {\bibfnamefont {T.}~\bibnamefont {Shibuya}},
		\bibinfo {author} {\bibfnamefont {J.}~\bibnamefont {Nakai}}, \bibinfo
		{author} {\bibfnamefont {K.}~\bibnamefont {Yoshiyasu}}, \bibinfo {author}
		{\bibfnamefont {Y.}~\bibnamefont {Sakai}}, \bibinfo {author} {\bibfnamefont
			{Y.}~\bibnamefont {Takano}}, \bibinfo {author} {\bibfnamefont
			{Y.}~\bibnamefont {Kamihara}},\ and\ \bibinfo {author} {\bibfnamefont
			{M.}~\bibnamefont {Matoba}},\ }\bibfield  {title} {\bibinfo {title} {The
			effect of simultaneous substitution on the electronic band structure and
			thermoelectric properties of {Se}-doped {Co$_3$SnInS$_2$} with the kagome
			lattice},\ }\href {https://doi.org/https://doi.org/10.1016/j.ssc.2014.09.006}
	{\bibfield  {journal} {\bibinfo  {journal} {Solid State Communications}\
		}\textbf {\bibinfo {volume} {199}},\ \bibinfo {pages} {56} (\bibinfo {year}
		{2014})}\BibitemShut {NoStop}%
	\bibitem [{\citenamefont {Li}\ \emph {et~al.}(2022)\citenamefont {Li},
		\citenamefont {Li}, \citenamefont {Lin}, \citenamefont {Guo}, \citenamefont
		{Sun}, \citenamefont {Chen}, \citenamefont {Luo}, \citenamefont {Yang},\ and\
		\citenamefont {Yuan}}]{Li2022Sb}%
	\BibitemOpen
	\bibfield  {author} {\bibinfo {author} {\bibfnamefont {Y.}~\bibnamefont
			{Li}}, \bibinfo {author} {\bibfnamefont {W.}~\bibnamefont {Li}}, \bibinfo
		{author} {\bibfnamefont {J.}~\bibnamefont {Lin}}, \bibinfo {author}
		{\bibfnamefont {Z.}~\bibnamefont {Guo}}, \bibinfo {author} {\bibfnamefont
			{F.}~\bibnamefont {Sun}}, \bibinfo {author} {\bibfnamefont {X.}~\bibnamefont
			{Chen}}, \bibinfo {author} {\bibfnamefont {Y.}~\bibnamefont {Luo}}, \bibinfo
		{author} {\bibfnamefont {J.}~\bibnamefont {Yang}},\ and\ \bibinfo {author}
		{\bibfnamefont {W.}~\bibnamefont {Yuan}},\ }\bibfield  {title} {\bibinfo
		{title} {Electron doping and physical properties in the ferromagnetic
			semimetal {Co$_3$Sn$_{2–x}$Sb$_x$S$_2$}},\ }\href
	{https://doi.org/10.1021/acs.jpcc.2c00236} {\bibfield  {journal} {\bibinfo
			{journal} {The Journal of Physical Chemistry C}\ }\textbf {\bibinfo {volume}
			{126}},\ \bibinfo {pages} {7230} (\bibinfo {year} {2022})}\BibitemShut
	{NoStop}%
	\bibitem [{\citenamefont {Yang}\ \emph {et~al.}(2020)\citenamefont {Yang},
		\citenamefont {You}, \citenamefont {Wang}, \citenamefont {Huang},
		\citenamefont {Xi}, \citenamefont {Xu}, \citenamefont {Cao}, \citenamefont
		{Tian}, \citenamefont {Xu}, \citenamefont {Dai},\ and\ \citenamefont
		{Li}}]{Yuke2020}%
	\BibitemOpen
	\bibfield  {author} {\bibinfo {author} {\bibfnamefont {H.}~\bibnamefont
			{Yang}}, \bibinfo {author} {\bibfnamefont {W.}~\bibnamefont {You}}, \bibinfo
		{author} {\bibfnamefont {J.}~\bibnamefont {Wang}}, \bibinfo {author}
		{\bibfnamefont {J.}~\bibnamefont {Huang}}, \bibinfo {author} {\bibfnamefont
			{C.}~\bibnamefont {Xi}}, \bibinfo {author} {\bibfnamefont {X.}~\bibnamefont
			{Xu}}, \bibinfo {author} {\bibfnamefont {C.}~\bibnamefont {Cao}}, \bibinfo
		{author} {\bibfnamefont {M.}~\bibnamefont {Tian}}, \bibinfo {author}
		{\bibfnamefont {Z.-A.}\ \bibnamefont {Xu}}, \bibinfo {author} {\bibfnamefont
			{J.}~\bibnamefont {Dai}},\ and\ \bibinfo {author} {\bibfnamefont
			{Y.}~\bibnamefont {Li}},\ }\bibfield  {title} {\bibinfo {title} {Giant
			anomalous {Nernst} effect in the magnetic {Weyl} semimetal
			{Co$_3$Sn$_2$S$_2$}},\ }\href
	{https://doi.org/10.1103/PhysRevMaterials.4.024202} {\bibfield  {journal}
		{\bibinfo  {journal} {Phys. Rev. Mater.}\ }\textbf {\bibinfo {volume} {4}},\
		\bibinfo {pages} {024202} (\bibinfo {year} {2020})}\BibitemShut {NoStop}%
	\bibitem [{\citenamefont {Guin}\ \emph {et~al.}(2019)\citenamefont {Guin},
		\citenamefont {Vir}, \citenamefont {Zhang}, \citenamefont {Kumar},
		\citenamefont {Watzman}, \citenamefont {Fu}, \citenamefont {Liu},
		\citenamefont {Manna}, \citenamefont {Schnelle}, \citenamefont {Gooth},
		\citenamefont {Shekhar}, \citenamefont {Sun},\ and\ \citenamefont
		{Felser}}]{Guin2019}%
	\BibitemOpen
	\bibfield  {author} {\bibinfo {author} {\bibfnamefont {S.~N.}\ \bibnamefont
			{Guin}}, \bibinfo {author} {\bibfnamefont {P.}~\bibnamefont {Vir}}, \bibinfo
		{author} {\bibfnamefont {Y.}~\bibnamefont {Zhang}}, \bibinfo {author}
		{\bibfnamefont {N.}~\bibnamefont {Kumar}}, \bibinfo {author} {\bibfnamefont
			{S.~J.}\ \bibnamefont {Watzman}}, \bibinfo {author} {\bibfnamefont
			{C.}~\bibnamefont {Fu}}, \bibinfo {author} {\bibfnamefont {E.}~\bibnamefont
			{Liu}}, \bibinfo {author} {\bibfnamefont {K.}~\bibnamefont {Manna}}, \bibinfo
		{author} {\bibfnamefont {W.}~\bibnamefont {Schnelle}}, \bibinfo {author}
		{\bibfnamefont {J.}~\bibnamefont {Gooth}}, \bibinfo {author} {\bibfnamefont
			{C.}~\bibnamefont {Shekhar}}, \bibinfo {author} {\bibfnamefont
			{Y.}~\bibnamefont {Sun}},\ and\ \bibinfo {author} {\bibfnamefont
			{C.}~\bibnamefont {Felser}},\ }\bibfield  {title} {\bibinfo {title}
		{Zero-field {Nernst} effect in a ferromagnetic kagome-lattice
			{Weyl}-semimetal {Co$_3$Sn$_2$S$_2$}},\ }\href
	{https://doi.org/10.1002/adma.201806622} {\bibfield  {journal} {\bibinfo
			{journal} {Advanced Materials}\ }\textbf {\bibinfo {volume} {31}},\ \bibinfo
		{pages} {1806622} (\bibinfo {year} {2019})}\BibitemShut {NoStop}%
	\bibitem [{\citenamefont {Ding}\ \emph {et~al.}(2019)\citenamefont {Ding},
		\citenamefont {Koo}, \citenamefont {Xu}, \citenamefont {Li}, \citenamefont
		{Lu}, \citenamefont {Zhao}, \citenamefont {Wang}, \citenamefont {Yin},
		\citenamefont {Lei}, \citenamefont {Yan}, \citenamefont {Zhu},\ and\
		\citenamefont {Behnia}}]{Ding2019}%
	\BibitemOpen
	\bibfield  {author} {\bibinfo {author} {\bibfnamefont {L.}~\bibnamefont
			{Ding}}, \bibinfo {author} {\bibfnamefont {J.}~\bibnamefont {Koo}}, \bibinfo
		{author} {\bibfnamefont {L.}~\bibnamefont {Xu}}, \bibinfo {author}
		{\bibfnamefont {X.}~\bibnamefont {Li}}, \bibinfo {author} {\bibfnamefont
			{X.}~\bibnamefont {Lu}}, \bibinfo {author} {\bibfnamefont {L.}~\bibnamefont
			{Zhao}}, \bibinfo {author} {\bibfnamefont {Q.}~\bibnamefont {Wang}}, \bibinfo
		{author} {\bibfnamefont {Q.}~\bibnamefont {Yin}}, \bibinfo {author}
		{\bibfnamefont {H.}~\bibnamefont {Lei}}, \bibinfo {author} {\bibfnamefont
			{B.}~\bibnamefont {Yan}}, \bibinfo {author} {\bibfnamefont {Z.}~\bibnamefont
			{Zhu}},\ and\ \bibinfo {author} {\bibfnamefont {K.}~\bibnamefont {Behnia}},\
	}\bibfield  {title} {\bibinfo {title} {Intrinsic anomalous {Nernst} effect
			amplified by disorder in a half-metallic semimetal},\ }\href
	{https://doi.org/10.1103/PhysRevX.9.041061} {\bibfield  {journal} {\bibinfo
			{journal} {Phys. Rev. X}\ }\textbf {\bibinfo {volume} {9}},\ \bibinfo {pages}
		{041061} (\bibinfo {year} {2019})}\BibitemShut {NoStop}%
	\bibitem [{\citenamefont {Kassem}\ \emph {et~al.}(2016)\citenamefont {Kassem},
		\citenamefont {Tabata}, \citenamefont {Waki},\ and\ \citenamefont
		{Nakamura}}]{Kassem2016}%
	\BibitemOpen
	\bibfield  {author} {\bibinfo {author} {\bibfnamefont {M.~A.}\ \bibnamefont
			{Kassem}}, \bibinfo {author} {\bibfnamefont {Y.}~\bibnamefont {Tabata}},
		\bibinfo {author} {\bibfnamefont {T.}~\bibnamefont {Waki}},\ and\ \bibinfo
		{author} {\bibfnamefont {H.}~\bibnamefont {Nakamura}},\ }\bibfield  {title}
	{\bibinfo {title} {Structure and magnetic properties of flux grown single
			crystals of {Co$_{3-x}$Fe$_x$Sn$_2$S$_2$} shandites},\ }\href
	{https://doi.org/https://doi.org/10.1016/j.jssc.2015.10.005} {\bibfield
		{journal} {\bibinfo  {journal} {Journal of Solid State Chemistry}\ }\textbf
		{\bibinfo {volume} {233}},\ \bibinfo {pages} {8} (\bibinfo {year}
		{2016})}\BibitemShut {NoStop}%
	\bibitem [{\citenamefont {Ding}\ \emph {et~al.}(2021)\citenamefont {Ding},
		\citenamefont {Koo}, \citenamefont {Yi}, \citenamefont {Xu}, \citenamefont
		{Zuo}, \citenamefont {Yang}, \citenamefont {Shi}, \citenamefont {Yan},
		\citenamefont {Behnia},\ and\ \citenamefont {Zhu}}]{Ding_2021}%
	\BibitemOpen
	\bibfield  {author} {\bibinfo {author} {\bibfnamefont {L.}~\bibnamefont
			{Ding}}, \bibinfo {author} {\bibfnamefont {J.}~\bibnamefont {Koo}}, \bibinfo
		{author} {\bibfnamefont {C.}~\bibnamefont {Yi}}, \bibinfo {author}
		{\bibfnamefont {L.}~\bibnamefont {Xu}}, \bibinfo {author} {\bibfnamefont
			{H.}~\bibnamefont {Zuo}}, \bibinfo {author} {\bibfnamefont {M.}~\bibnamefont
			{Yang}}, \bibinfo {author} {\bibfnamefont {Y.}~\bibnamefont {Shi}}, \bibinfo
		{author} {\bibfnamefont {B.}~\bibnamefont {Yan}}, \bibinfo {author}
		{\bibfnamefont {K.}~\bibnamefont {Behnia}},\ and\ \bibinfo {author}
		{\bibfnamefont {Z.}~\bibnamefont {Zhu}},\ }\bibfield  {title} {\bibinfo
		{title} {Quantum oscillations, magnetic breakdown and thermal {Hall} effect
			in {Co$_3$Sn$_2$S$_2$}},\ }\href {https://doi.org/10.1088/1361-6463/ac1c2b}
	{\bibfield  {journal} {\bibinfo  {journal} {Journal of Physics D: Applied
				Physics}\ }\textbf {\bibinfo {volume} {54}},\ \bibinfo {pages} {454003}
		(\bibinfo {year} {2021})}\BibitemShut {NoStop}%
	\bibitem [{\citenamefont {Pasupathy}\ \emph {et~al.}(2004)\citenamefont
		{Pasupathy}, \citenamefont {Bialczak}, \citenamefont {Martinek},
		\citenamefont {Grose}, \citenamefont {Donev}, \citenamefont {McEuen},\ and\
		\citenamefont {Ralph}}]{Pasupathy2004}%
	\BibitemOpen
	\bibfield  {author} {\bibinfo {author} {\bibfnamefont {A.~N.}\ \bibnamefont
			{Pasupathy}}, \bibinfo {author} {\bibfnamefont {R.~C.}\ \bibnamefont
			{Bialczak}}, \bibinfo {author} {\bibfnamefont {J.}~\bibnamefont {Martinek}},
		\bibinfo {author} {\bibfnamefont {J.~E.}\ \bibnamefont {Grose}}, \bibinfo
		{author} {\bibfnamefont {L.~A.~K.}\ \bibnamefont {Donev}}, \bibinfo {author}
		{\bibfnamefont {P.~L.}\ \bibnamefont {McEuen}},\ and\ \bibinfo {author}
		{\bibfnamefont {D.~C.}\ \bibnamefont {Ralph}},\ }\bibfield  {title} {\bibinfo
		{title} {The {Kondo} effect in the presence of ferromagnetism},\ }\href
	{https://doi.org/10.1126/science.1102068} {\bibfield  {journal} {\bibinfo
			{journal} {Science}\ }\textbf {\bibinfo {volume} {306}},\ \bibinfo {pages}
		{86} (\bibinfo {year} {2004})}\BibitemShut {NoStop}%
	\bibitem [{\citenamefont {Krellner}\ \emph {et~al.}(2007)\citenamefont
		{Krellner}, \citenamefont {Kini}, \citenamefont {Brüning}, \citenamefont
		{Koch}, \citenamefont {Rosner}, \citenamefont {Nicklas}, \citenamefont
		{Baenitz},\ and\ \citenamefont {Geibel}}]{Krellner2007}%
	\BibitemOpen
	\bibfield  {author} {\bibinfo {author} {\bibfnamefont {C.}~\bibnamefont
			{Krellner}}, \bibinfo {author} {\bibfnamefont {N.~S.}\ \bibnamefont {Kini}},
		\bibinfo {author} {\bibfnamefont {E.~M.}\ \bibnamefont {Brüning}}, \bibinfo
		{author} {\bibfnamefont {K.}~\bibnamefont {Koch}}, \bibinfo {author}
		{\bibfnamefont {H.}~\bibnamefont {Rosner}}, \bibinfo {author} {\bibfnamefont
			{M.}~\bibnamefont {Nicklas}}, \bibinfo {author} {\bibfnamefont
			{M.}~\bibnamefont {Baenitz}},\ and\ \bibinfo {author} {\bibfnamefont
			{C.}~\bibnamefont {Geibel}},\ }\bibfield  {title} {\bibinfo {title}
		{{CeRuPO}: A rare example of a ferromagnetic {Kondo} lattice},\ }\href
	{https://doi.org/10.1103/PhysRevB.76.104418} {\bibfield  {journal} {\bibinfo
			{journal} {Physical Review B}\ }\textbf {\bibinfo {volume} {76}},\ \bibinfo
		{pages} {104418} (\bibinfo {year} {2007})}\BibitemShut {NoStop}%
	\bibitem [{\citenamefont {Tursina}\ \emph {et~al.}(2018)\citenamefont
		{Tursina}, \citenamefont {Khamitcaeva}, \citenamefont {Gnida},\ and\
		\citenamefont {Kaczorowski}}]{Tursina2018}%
	\BibitemOpen
	\bibfield  {author} {\bibinfo {author} {\bibfnamefont {A.}~\bibnamefont
			{Tursina}}, \bibinfo {author} {\bibfnamefont {E.}~\bibnamefont
			{Khamitcaeva}}, \bibinfo {author} {\bibfnamefont {D.}~\bibnamefont {Gnida}},\
		and\ \bibinfo {author} {\bibfnamefont {D.}~\bibnamefont {Kaczorowski}},\
	}\bibfield  {title} {\bibinfo {title} {{CePd$_2$Al$_8$} – a ferromagnetic
			{Kondo} lattice with new type of crystal structure},\ }\href
	{https://doi.org/https://doi.org/10.1016/j.jallcom.2017.10.031} {\bibfield
		{journal} {\bibinfo  {journal} {Journal of Alloys and Compounds}\ }\textbf
		{\bibinfo {volume} {731}},\ \bibinfo {pages} {229} (\bibinfo {year}
		{2018})}\BibitemShut {NoStop}%
	\bibitem [{\citenamefont {Das}\ and\ \citenamefont
		{Kaczorowski}(2019)}]{Das2019}%
	\BibitemOpen
	\bibfield  {author} {\bibinfo {author} {\bibfnamefont {D.}~\bibnamefont
			{Das}}\ and\ \bibinfo {author} {\bibfnamefont {D.}~\bibnamefont
			{Kaczorowski}},\ }\bibfield  {title} {\bibinfo {title} {Ferromagnetic {Kondo}
			lattice behavior in {Ce$_{11}$Pd$_4$In$_9$}},\ }\href
	{https://doi.org/https://doi.org/10.1016/j.jmmm.2018.09.104} {\bibfield
		{journal} {\bibinfo  {journal} {Journal of Magnetism and Magnetic Materials}\
		}\textbf {\bibinfo {volume} {471}},\ \bibinfo {pages} {315} (\bibinfo {year}
		{2019})}\BibitemShut {NoStop}%
	\bibitem [{\citenamefont {Kassem}\ \emph {et~al.}(2020)\citenamefont {Kassem},
		\citenamefont {Tabata}, \citenamefont {Waki},\ and\ \citenamefont
		{Nakamura}}]{Kassem2021}%
	\BibitemOpen
	\bibfield  {author} {\bibinfo {author} {\bibfnamefont {M.~A.}\ \bibnamefont
			{Kassem}}, \bibinfo {author} {\bibfnamefont {Y.}~\bibnamefont {Tabata}},
		\bibinfo {author} {\bibfnamefont {T.}~\bibnamefont {Waki}},\ and\ \bibinfo
		{author} {\bibfnamefont {H.}~\bibnamefont {Nakamura}},\ }\bibfield  {title}
	{\bibinfo {title} {Unconventional critical behaviors at the magnetic phase
			transition of {Co$_3$Sn$_2$S$_2$} kagomé ferromagnet},\ }\href
	{https://doi.org/10.1088/1361-648X/abaf94} {\bibfield  {journal} {\bibinfo
			{journal} {Journal of Physics: Condensed Matter}\ }\textbf {\bibinfo {volume}
			{33}},\ \bibinfo {pages} {015801} (\bibinfo {year} {2020})}\BibitemShut
	{NoStop}%
	\bibitem [{\citenamefont {Behnia}\ \emph {et~al.}(2004)\citenamefont {Behnia},
		\citenamefont {Jaccard},\ and\ \citenamefont {Flouquet}}]{Kamran2004}%
	\BibitemOpen
	\bibfield  {author} {\bibinfo {author} {\bibfnamefont {K.}~\bibnamefont
			{Behnia}}, \bibinfo {author} {\bibfnamefont {D.}~\bibnamefont {Jaccard}},\
		and\ \bibinfo {author} {\bibfnamefont {J.}~\bibnamefont {Flouquet}},\
	}\bibfield  {title} {\bibinfo {title} {On the thermoelectricity of correlated
			electrons in the zero-temperature limit},\ }\href
	{https://doi.org/10.1088/0953-8984/16/28/037} {\bibfield  {journal} {\bibinfo
			{journal} {Journal of Physics: Condensed Matter}\ }\textbf {\bibinfo
			{volume} {16}},\ \bibinfo {pages} {5187} (\bibinfo {year}
		{2004})}\BibitemShut {NoStop}%
	\bibitem [{\citenamefont {Hu}\ \emph {et~al.}(2022)\citenamefont {Hu},
		\citenamefont {Kan}, \citenamefont {Chen}, \citenamefont {Zheng},\ and\
		\citenamefont {Ma}}]{Hu2022specific}%
	\BibitemOpen
	\bibfield  {author} {\bibinfo {author} {\bibfnamefont {J.}~\bibnamefont
			{Hu}}, \bibinfo {author} {\bibfnamefont {X.}~\bibnamefont {Kan}}, \bibinfo
		{author} {\bibfnamefont {Z.}~\bibnamefont {Chen}}, \bibinfo {author}
		{\bibfnamefont {G.}~\bibnamefont {Zheng}},\ and\ \bibinfo {author}
		{\bibfnamefont {Y.}~\bibnamefont {Ma}},\ }\bibfield  {title} {\bibinfo
		{title} {The magnetic, thermal transport properties, magnetothermal effect
			and critical behavior of {Co$_3$Sn$_2$S$_2$} single crystal},\ }\href
	{https://doi.org/https://doi.org/10.1111/jace.18465} {\bibfield  {journal}
		{\bibinfo  {journal} {Journal of the American Ceramic Society}\ }\textbf
		{\bibinfo {volume} {105}},\ \bibinfo {pages} {4827} (\bibinfo {year}
		{2022})}\BibitemShut {NoStop}%
	\bibitem [{\citenamefont {Robinson}(1967)}]{Robinson1967}%
	\BibitemOpen
	\bibfield  {author} {\bibinfo {author} {\bibfnamefont {J.~E.}\ \bibnamefont
			{Robinson}},\ }\bibfield  {title} {\bibinfo {title} {Thermoelectric power in
			the nearly-free-electron model},\ }\href@noop {} {\bibfield  {journal}
		{\bibinfo  {journal} {Phys. Rev.}\ }\textbf {\bibinfo {volume} {161}},\
		\bibinfo {pages} {533} (\bibinfo {year} {1967})}\BibitemShut {NoStop}%
	\bibitem [{\citenamefont {Asaba}\ \emph {et~al.}(2021)\citenamefont {Asaba},
		\citenamefont {Ivanov}, \citenamefont {Thomas}, \citenamefont {Savrasov},
		\citenamefont {Thompson}, \citenamefont {Bauer},\ and\ \citenamefont
		{Ronning}}]{asaba2021}%
	\BibitemOpen
	\bibfield  {author} {\bibinfo {author} {\bibfnamefont {T.}~\bibnamefont
			{Asaba}}, \bibinfo {author} {\bibfnamefont {V.}~\bibnamefont {Ivanov}},
		\bibinfo {author} {\bibfnamefont {S.~M.}\ \bibnamefont {Thomas}}, \bibinfo
		{author} {\bibfnamefont {S.~Y.}\ \bibnamefont {Savrasov}}, \bibinfo {author}
		{\bibfnamefont {J.~D.}\ \bibnamefont {Thompson}}, \bibinfo {author}
		{\bibfnamefont {E.~D.}\ \bibnamefont {Bauer}},\ and\ \bibinfo {author}
		{\bibfnamefont {F.}~\bibnamefont {Ronning}},\ }\bibfield  {title} {\bibinfo
		{title} {Colossal anomalous {Nernst} effect in a correlated
			noncentrosymmetric kagome ferromagnet},\ }\href
	{https://doi.org/10.1126/sciadv.abf1467} {\bibfield  {journal} {\bibinfo
			{journal} {Science Advances}\ }\textbf {\bibinfo {volume} {7}},\ \bibinfo
		{pages} {eabf1467} (\bibinfo {year} {2021})}\BibitemShut {NoStop}%
\end{thebibliography}
\end{document}